\documentclass[onecolumn,floatfix,amsmath,amssymb]{revtex4-1}
\usepackage{amsmath,amsfonts,amsthm,bm}
\usepackage{graphicx}
\usepackage{dcolumn} 
\usepackage{bm}      
\usepackage[usenames,dvipsnames]{color}
\usepackage{ulem}    

\usepackage{hyperref}
\newcommand{\MYhref}[3][red]{\href{#2}{\color{#1}{#3}}}
\begin{document}

\title{On the limitations of some popular numerical models of flagellated microswimmers: importance of long-range forces and flagellum waveform}

\author{
C. Rorai$^{1}$, M. Zaitsev$^{2}$ and S. Karabasov$^{1}$}

\address{$^{1}$School of Engineering and Materials Science, Queen Mary University of London, Mile End Road,
London E1 4NS;\\
$^{2}$Nuclear Safety Institute, ul. Bolshaja Tulskaja, 52, 115191, Moscow.}




\begin{abstract}
For a sperm cell-like flagellated swimmer in an unbounded domain, several numerical models of different fidelity are considered based on the Stokes flow approximation. The models include a Regularised Stokeslet Method and a 3D Finite Element Method, which serve as the benchmark solutions for several approximate models considered. The latter include the Resistive Force Theory versions of Lighthill and Gray and Hancock as well as a simplified approximation based on computing the hydrodynamic forces exerted on the head and the flagellum separately. It is shown how none of the simplified models is robust enough with regards to predicting the effect of the swimmer head shape change on the swimmer dynamics. For a range of swimmer motions considered, the resulting solutions for the swimmer force and  velocities are analysed and the applicability of the Stokes model for the swimmers in question is probed. 
\end{abstract}


\maketitle 
\noindent
\section{Introduction}\label{sec:intro}

Flagellated microswimmers are cells or micrometer-size robots that swim by moving appendages called flagella. Bacteria flagella appear as helical filaments rigidly rotated by a motor complex attached to the cell wall, eukaryotic flagella, instead, move by propagating sinusoidal waves in a whip-like fashion. The reason of this difference lies in the specific structure of eukaryotic 
 cilia and flagella: the axoneme \cite{ElgetiReview}. 
The ability to model flagellated microswimmers mathematically and numerically is relevant to a variety of applications in the fields of biology, medicine, medical diagnostic and engineering \cite{Stanton15}. Beside improving our understanding of the physical phenomenon, accurate models can inform the design of effective microfluidic devices to sort microswimmers by motility \cite{Lambert10}, \cite{Denissenko12}, \cite{Nosrati17} or suggest the design of efficient artificial microswimmers \cite{Williams14}.

In this study we are specifically concerned with the hydrodynamical modeling of sperm-cell swimmers. The dimensionless ratio between the inertia and viscous forces acting on these cells, namely, the Reynolds number, is of the order of Re $= UL/\nu \approx10^{-2}$, while the ratio between the characteristic viscous time scale and the time scale representing the rate of deformation of the swimmer body, {\it i.e.} the frequency Reynolds number, is of the order of Re$_{\omega}$ $=L^2\omega/\nu\approx 10^{-1}$. Here $L\approx 50 \mu m$ and $U\approx 2\cdot 10^{-4}m/s$ are the characteristic length and velocity of the swimmer, $\omega\approx 32 s^{-1}$ is the beating frequency of the flagellum and $\nu=10^{-6} m^2/s$ is the kinematic viscosity of water. For Re and Re$_{\omega}<<1$, the flow field is typically computed by integrating the Stokes equations in a time-independent zero-Reynolds number framework, see for example \cite{Ishimoto14} and \cite{Ishimoto16}. It can be noted,  however, that neither of the two standard Reynolds number definitions include any scale associated with a change of the swimmer shape such as a characteristic wavelength of the flagellum motion that is not uniquely defined by the beating frequency in case of a sperm-cell swimmer. Thus, in the latter case, the applicability of the common criteria of ignoring the unsteady and inertial effects based on the two Reynolds numbers being of o(1) can be debated. It can be noted that classical studies \cite{Johnson79} avoid this controversy by considering simplified flagellated swimmer models, which, for example, cannot capture the hydrodynamically important details of the flagellum waveform near the open ends, and assume that the wavelength of the swimmer's
motion is equivalent to its linear size. Although for a simple sperm cell swimming  along a straight trajectory  the wavelength is more or less equal to the length of the flagellum, in general, for more complex trajectories of the sperm cell that lncludes sharp turns or other organisms and artificial swimmers this is not the case.

The Resistive Force Theory (RFT) \cite{Lighthill} is a further simplified model applicable to the motion of slender bodies of which beating flagella is a good example. The RFT neglects the long range hydrodynamical interactions and evaluates the viscous forces exerted on the immersed body as a function of the local velocities only. This model presents many advantages: it has a low computational cost when compared with the numerical integration of the Stokes equations allowing for proof of concept calculations \cite{Alouges13, Montino15}, and it is simple enough to serve as a starting point for further analytical derivations \cite{Lauga13, Man16}, yet its accuracy is debated. Early works discussed the best choice for the model parameters, namely the normal and tangential hydrodynamical friction coefficients, different proposals were put forward by Lighthill \cite{Lighthill75} and Gray and Hancock \cite{Gray55}. Recent experimental tests showed that either choices badly capture the behavior of helical flagella for the range of shapes present in nature \cite{Rodenborn13, Jung07}. Other studies calibrated the parameters to match experimental observations \cite{Friedrich10} or the results obtained by integrating Stokes \cite{Zaitsev}. 
The necessity of calibrating the model versus experimental observations or the results of more sophisticated models highlights the unsuitability of the RFT approximation for applications outside the range of calibration.  

The RFT is applied to the flagellum, the contribution of the approximately ellipsoidal cell-body or ``head'' is evaluated through analytical expressions \cite{Chwang75} and added to the flagellum contribution to compute the entire cell dynamics. On this same line one may represent the flagellum by a model of choice and still approach the problem by separately studying the cell body and flagellum dynamics before simply summing their contributions \cite{Giuliani17}. This procedure is naturally embodied in the RFT and is for example implied by studies that look for the optimal flagellum shape neglecting to include the head in the calculations {\it e.g} \cite{Lighthill, Lauga13}. 

In this paper we study the motion of a single sperm-cell in an infinite domain to address three issues: 
({\it i}) the accuracy of approximating swimming as the linear superposition of the dynamics of separate body-parts (head plus 
flagellum) versus a full-body description, ({\it ii}) the
accuracy of modelling swimming with the Resistive Force Theory (RFT), which is one of the possible approximations based on the above superposition that ignores long-range hydrodynamical interactions, and ({\it iii}) the validity of the quasi-steady and
inertia-less assumption for swimming at the micro-scale,
where the quasi-steady hypothesis entails assuming that
the flow instantaneously adapts to the body deformations. 

To address the first two issues we contrast the results obtained by applying the simplified approaches and the full hydrodynamical model, section \ref{RFT} and \ref{simplified}. In particular, we compare the swimming velocities, trajectory and force distribution on the swimmer body. We then study locomotion of swimmers with different head shapes, section \ref{head}, and show that the simplified approaches fail to identify the most hydrodynamically efficient swimmer. To tackle the third point we calculate the propulsive matrix, that is the matrix that relates the forces generated by the moving flagellum to the rigid-body velocities of the swimmers, and we analyze its eigenvalues and eigenvectors to identify a criterion that establishes when the inertia-less quasi-steady hypothesis is valid, section \ref{eigenvalues}.


It can be noted that the investigation of the applicability limits of RFT has been a popular topic since the 1970s. For example, in the already mentioned study of Johnson \& Brokaw \cite{Johnson79} it was shown that RFT is satisfactory for use in analysis of mechanisms for the control of flagellar bending in the analytical framework of Brokaw \cite{Brokaw72} that simulates the behavior of a spermatozoa flagellum by an active shear system controlled by the curvature of the flagellum. In that work a good agreement in comparison with the Slender Body Theory (SBT) developed by Johnson \cite{Johnson80} was reported. The SBT can be viewed as a simplified version of RSM where the solution of the Stokes flow problem resulting from the motion of a slender body is developed using the singularity method assuming that the cross-sectional area varies slowly along its length. In the original SBT framework, the cell (head) effect on hydrodynamics is modelled separately from the flagellum using the analytical Stokes sphere solution. For certain idealised flagellum approximations, such as representing the flagellum body by a thin rigid helix that rotates with a constant angular velocity, the SBT model was shown to be consistent with the Boundary Element Method \cite{Ramia93} at the level of computational resolution affordable at the time. 
However, to the best knowledge of the authors, a systematic comparison of RFT, a semi-analytical integral method, and a direct solution of the Stokes equation in case of a realistic, flexible flagellum waveform such as the one described in \cite{Alouges13} has not been performed yet. 
SBT is not considered separately in this publication which focuses on a more sophisticated integral method, i.e. RSM, which can include the effect of a finite thickness of the flagellum and it is not limited to analytical approximations to account for the hydrodynamic contribution of the head. The RSM solutions will be validated in comparison with Finite Element Method at an order of magnitude higher resolution of the swimmer's body including the head in comparison with the previous studies \cite{Ramia93} and then compared with RFT.

The paper is organized as follows: in section \ref{sec:method} we introduce the numerical methods used to simulate the flagellated swimmer motion. The Regularized Stokeslet Method (RSM) is presented in section \ref{RSM}, the swimming
problem details in the context of the RSM are discussed in \ref{Theswimmingproblem}, while in \ref{subsec:parameters} the geometry and beating movement of the flagellum are defined. 
The application of a Finite Element Method for the same microswimmer problem is introduced in section \ref{subsec:FEM}.
In section \ref{sec:numerics} we report the numerical results including a validation of the RSM versus the FEM code, a modal analysis of the swimming velocities and a visualization of the flow field induced by the swimmer (section \ref{Flowfield}). 
We summarize the main results in section \ref{sec:conclusions}.

\section{Mathematical Models} \label{sec:method}
\subsection{Regularized Stokeslet Method}\label{RSM}
The fundamental solution of the incompressible forced Stokes equation 
 \begin{equation}
-\nabla p+\mu \nabla^2\bf{u}+\bf{f}\delta(\bf{x}-\bf{y}) = 0,
\label{forcedSt}  \end{equation}
 \begin{equation}
\nabla \cdot \bf{u} = 0,
\label{forcedSt}  \end{equation}
for a point force ${\bf f}$ acting on ${\bf y}$ in an unbounded domain  is the {\it Stokeslet} $\mathcal{J}({\bf r})$:
\begin{equation}
{\bf u}({\bf x})={\bf f}\cdot \mathcal{J}({\bf r})=\frac{{\bf f}}{8\pi\mu}\cdot\left[\frac{\mathcal{I}}{r}+\frac{{\bf r}{\bf r}^T}{r^3}\right],
\end{equation}
where $r=|\bf{x}-\bf{y}|$, ${\bf r}=(\bf{x}-\bf{y})$, $\mathcal{I}$ is the identity matrix, $\mu$ is the dynamic viscosity of the fluid, and $\delta$ is the $\delta$-function.

Since the Stokes equation is linear, the flow field generated by an immersed body with a deforming boundary $\mathcal{S}(t)$ can be represented through a continuous distribution of Stokeslets \cite{Pozrikidisbook}:
\begin{equation}
{\bf u}({\bf x}, t)=\int_{\mathcal{S}(t)}{\bf f}({\bf y})\cdot \mathcal{J({\bf r})}d\mathcal{S}_{{\bf y}}.\label{BI}
\end{equation}
For complex geometries, as is the case of flagellated micro-swimmers, the integral is computed numerically by discretizing the immersed surface: for $n$, the points on which the velocity is evaluated, $m$, the grid points on the surface of the immersed body, and $A_m$ the quadrature weights,
\begin{equation}
{\bf u}_{n}=\frac{1}{8\pi\mu}\sum_{m} \left(\frac{\mathcal{I}}{|{\bf r}_{nm}|}+\frac{{\bf r}_{nm} {\bf r}_{nm}^{T}}{|{\bf r}_{nm}|^3}\right)\cdot {\bf f}_{m} A_m, \label{BIdiscr}
\end{equation}
where ${\bf r}_{nm}={\bf x}_{n}-{\bf x}_{m}$ is the radius vector from point $n$ to $m$.
In our calculations the RSM grid is built on the surface of the swimmer head and on the cylindrical surface of the flagellum of radius $F_r$, see section \ref{subsec:parameters} for details on the swimmer geometry.
 
The Stokeslets are singular kernels, their singularity can be dealt with  by replacing the point force with an approximate point force with local support. In practice, as proposed by Cortez \cite{cortezFauci_regusto}, ${\bf f}\delta({\bf r})$ can be replaced by ${\bf f}\phi_{\epsilon}({\bf r})$ 
\begin{equation}
\phi_{\epsilon}({\bf r})=\frac{15\epsilon^4}{8\pi(r^2+\epsilon)^{7/2}},
\end{equation}
yielding the {\it regularized} Stokeslet: 
\begin{equation}
\mathcal{J}^{\epsilon}_{ij}=\delta_{ij}\frac{r^2+2\epsilon^2}{(r^2+\epsilon^2)^{3/2}}+\frac{(x_{n,i}-x_{m,i})(x_{n,j}-x_{m,j})}{(r^2+\epsilon^2)^{3/2}},
\end{equation}
for which ~97\% of the force is within a radius $\epsilon$ \cite{cortezFauci_regusto}, where  $\epsilon$ is 
the regularization parameter that requires calibration. By running some tests for the flow past an ellipsoid (see section \ref{Theswimmingproblem}) we have found, consistently with \cite{Rodenborn13}, that our numerical results minimize the error when $\epsilon$ is between one third and one half of the grid spacing. 

\subsection{The swimming problem}\label{Theswimmingproblem}

\begin{figure} \centering
\resizebox{7.5cm}{!}{\includegraphics{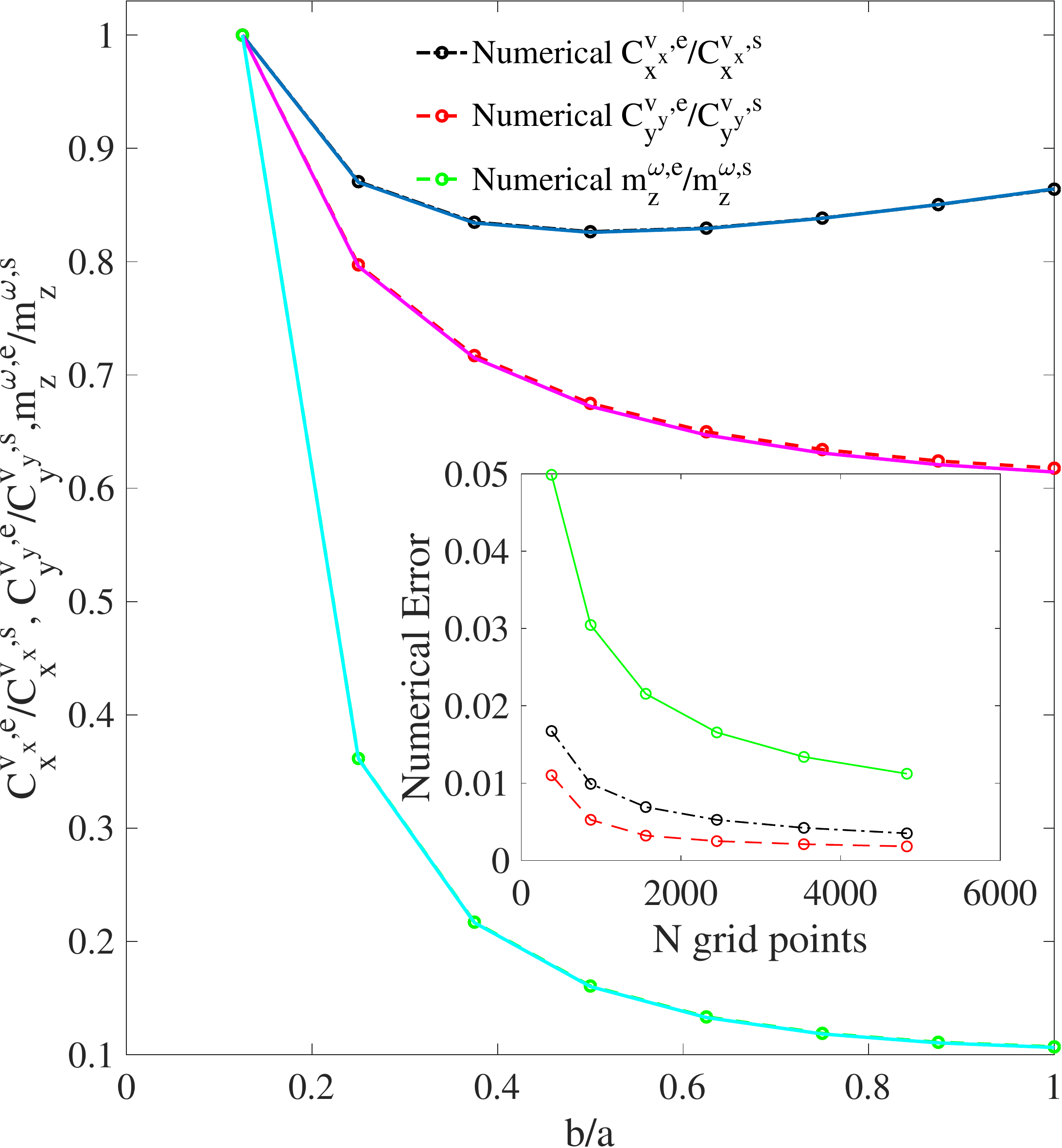}}
\caption{\label{NumericalErrorHead}
Ratio between $C_x^{v_x}$ for a prolate ellipsoid, namely $C_x^{v_x,e}$, and $C_x^{v_x}$ for a sphere of equal volume, namely $C_x^{v_x,s}$ (black dashed-dotted line with circles and solid blue line), similarly, the ratio $C_y^{v_y,e}/C_y^{v_y,s}$ (red dashed line with circles and magenta solid line) and $m_z^{\omega,e}/m_z^{\omega,s}$ (green dashed line with circles and cyan solid line). The curves are rescaled by their maximum value to facilitate a direct comparison between them. The solid lines correspond to the analytical solution of eq. (\ref{cxvxhe})-(\ref{mzomegae}), while the dashed lines with empty circles represent the numerical solution for an ellipsoid discretized by 4694 points.  The inset shows the numerical error for these same quantities as specified by eq. (\ref{erroroncxcz}) and as a function of the number of points the surface is discretized into.}
\end{figure}

The velocity of a swimmer in Stokes flow can be decomposed into a rigid-body translation $v_j$, a rigid rotation $\omega_j$, and the body deformation (head and flagellum) $u_j^{BC}({\bf x})$. For convenience, we express these velocities in the frame of reference of the swimmer. The head is non-motile ($u_j^{BC}=0$), while the flagellum moves with the beating motion introduced in section \ref{subsec:parameters}. We consider the case of a flagellum beating on the $z=0$ plane.

In the context of the Regularized Stokeslet Method the swimming problem is solved by inverting the system 
\begin{equation}
u_j^{BC}({\bf x})=\frac{1}{8 \pi \mu}\sum_{n=1}^{N}\sum_{i=1}^{3}\mathcal{J}^{\epsilon}_{ij}({\bf x},{\bf x}_n)f_{n,i}A_n-v_j-{\bm \omega}\times{\bf x}\cdot e_j
\label{full_system}
\end{equation}
with constraints
\begin{align}
\label{inertialess1}
f_{n,i}A_n&=0,\\
F_{n,i} x_{n,j} \varepsilon_{ijk}&=0, 
\label{inertialess}
\end{align}
to find the forces $F_{n,i}$ and the swimming velocities $v_j$ and $\omega_j$. Here $F_{n,i}=f_{n,i}A_n$, for $n=1,2 ... N$ and in equations (\ref{inertialess1})-(\ref{inertialess}) the summation over the repeated index convention is adopted, with $i=1,2,3$ as well as $j$ and $k$. The conditions (\ref{inertialess1})-(\ref{inertialess}) derive from the fact that forces and torques need to balance exactly since inertia is absent.  

An alternative but equivalent approach consists in computing the propulsive matrix coefficients $C_x$, $C_y$, $m_z$ 
\begin{equation}
\left[
\begin{array}{ccc}
C_x^{v_x} & C_x^{v_y} & C_x^{\omega} \\
C_y^{v_x} & C_y^{v_y} & C_y^{\omega} \\
m_z^{v_x} & m_z^{v_y} & m_z^{\omega}
\end{array} 
\right]\label{prop_matrix}
\end{equation}
and solving the system  
\begin{align}\label{systemswimming1}
C_x^{v_x}v_x+C_x^{v_y}v_y+C_x^{\omega}\omega&=-F_x^B\\
\label{systemswimming2}
C_y^{v_x}v_x+C_y^{v_y}v_y+C_y^{\omega}\omega&=-F_y^B\\
m_z^{v_x}v_x+m_z^{v_y}v_y+m_z^{\omega}\omega&=-T_z^B,
\label{systemswimming3}
\end{align}
for $v_x$, $v_y$ and $\omega$.
The coefficients $C$ and $m$ are computed by solving the Stokes equation separately for an arbitrary (unitary for convenience) rigid translation of the swimmer body and an arbitrary solid body rotation. They correspond to the surface integral on the swimmer body of the $x$, eq. (\ref{systemswimming1}), and $y$, eq. (\ref{systemswimming2}), component of the force density $f_i$, and the $z$, eq. (\ref{systemswimming3}), component of the torque density for, respectively, a unitary $v_x$, $v_y$ and $\omega$. The known terms $F_x^B$, $F_y^B$, $T_z^B$ are the integrals of the forces and torque due to the flagellum beating only. This approach requires solving equation (\ref{full_system}) four times (for an arbitrary $v_x$, $v_y$, $\omega$ and for the flagellum beating) but has the advantage of producing better conditioned matrices. We recall that a direct consequence of the reciprocal theorem is that the resistance matrix (\ref{prop_matrix}) is symmetric.  

If we approximate the swimming problem by treating the flagellum and the head separately, the propulsive matrix for the frame of reference located on the head centroid becomes:
\begin{equation}
\left[ 
\begin{array}{lll}
(C_x^{v_x,t}+C_x^{v_x,h}) & C_x^{v_y} & C_x^{\omega,t} \\
C_y^{v_x} & (C_y^{v_y,t}+C_y^{v_y,h}) & C_y^{\omega,t} \\
m_z^{v_x,t} & m_z^{v_y,t}& (m_z^{\omega,t}+m_z^{\omega,h})
\end{array}
\right]
\label{tail+head}\end{equation}
where the superscript $t$ and $h$ stand for the tail (flagellum) contribution and the head contribution. We will next refer to this approach as the head+tail (H+T) model.

If the head is spherical $C_x^{v_x,h}=C_y^{v_y,h}=6\pi\mu L_{head}/2$ and $m_z^{\omega,h}=8\pi\mu(L_{head}/2)^3$. 
Expressions for $C_x^{v_x}$ $C_y^{v_y}$ and $m_z^{\omega}$ for a prolate ellipsoid with $b=c$ as the minor semi-axes and $a$ as the major semi-axis are derived in \cite{Chwang75}:
\begin{equation}
C_x^{v_x}= 6\pi\mu a\frac{8}{3}e^3\left[-2e+(1+e^2)\log{\frac{1+e}{1-e}}\right]^{-1}
\label{cxvxhe}\end{equation}
\begin{equation}
C_y^{v_y}= 6\pi\mu a\frac{16}{3}e^3\left[2e+(3e^2-1)\log{\frac{1+e}{1-e}}\right]^{-1},
\label{cyvyhe}\end{equation}
\begin{equation}
m_z^{\omega}= 8\pi\mu a b^2 \frac{4}{3}e^3\left(\frac{2-e^2}{1-e^2}\right)\left[-2e+(e^2+1)\log{\frac{1+e}{1-e}}\right]^{-1},
\label{mzomegae}\end{equation}
where $e=\sqrt{1-(b/a)^2}$ is the eccentricity, and $m_z^{\omega}$ is calculated for a rotation about a minor axis.

To test the accuracy of the Regularized Stokeslet Method we study the flow past a prolate ellipsoid for $N_e$ different stretching ratios, $b/a$, in the range $0.125<b/a<1$.
A comparison between the analytical result, {\it i.e.} expressions (\ref{cxvxhe})-(\ref{mzomegae}), and the numerical result obtained with the Regularized Stokeslet Method is shown in Fig. \ref{NumericalErrorHead}. After calibration, the regularization parameter $\epsilon$ is chosen to be $\epsilon=0.5\sqrt{\mathcal{S}/N}=0.5\sqrt{\mathcal{A}}$, where $\mathcal{S}$ is the surface of the ellipsoid and $N$ is the number of grid points the ellipsoid is represented by.
 The numerical error, plotted in the inset of Fig. \ref{NumericalErrorHead}, is computed as
\begin{equation}
\text{Error}=\frac{1}{N_e}\sum_i\frac{|C_{numerical}^{i}-C_{analytical}^{i}|}{C_{analytical}^{i}},
\label{erroroncxcz}\end{equation}
where $C$ stands for the coefficient $C_x^{v_x}$, $C_y^{v_y}$ or $m_z^{\omega}$ for a prolate ellipsoid divided by the same coefficient for a sphere of equal volume. Note that for a given volume, the surface area increases as the aspect ratio decreases. 

We have compared the results for two different distributions of the grid points: the case of equally spaced points on the surface of the ellipsoid and an the case of an uneven grid (a discretization based on spherical coordinates with points located at equal azimuthal and polar angle intervals). We finally chose to adopt the latter since the results are not very sensitive to the type of grid once the resolution is large enough.

\subsection{Swimming parameters}\label{subsec:parameters}

\begin{figure}
\begin{tabular}{lc}
\resizebox{6.0cm}{!}{\includegraphics{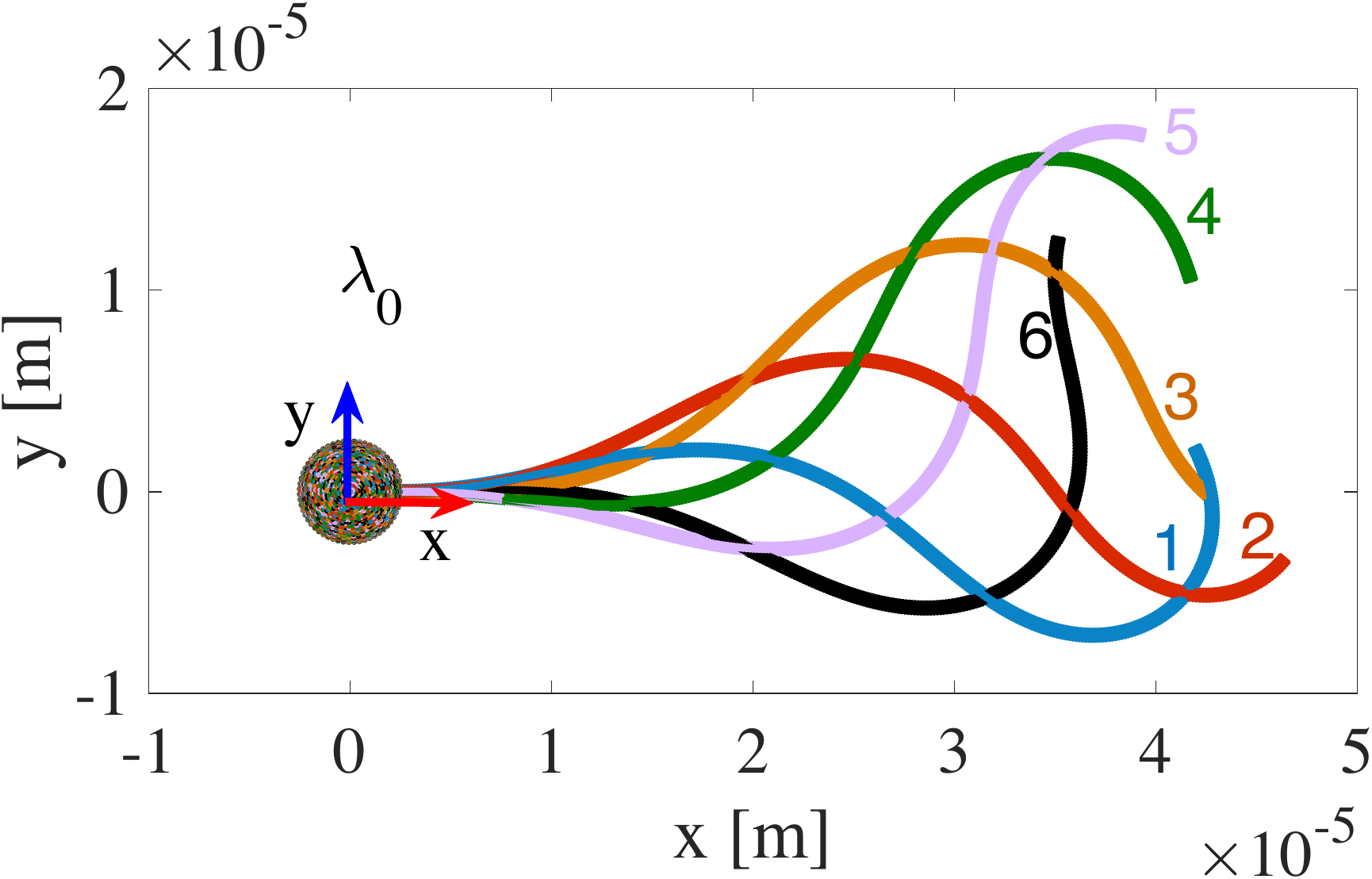}}
& \resizebox{6.0cm}{!}{\includegraphics{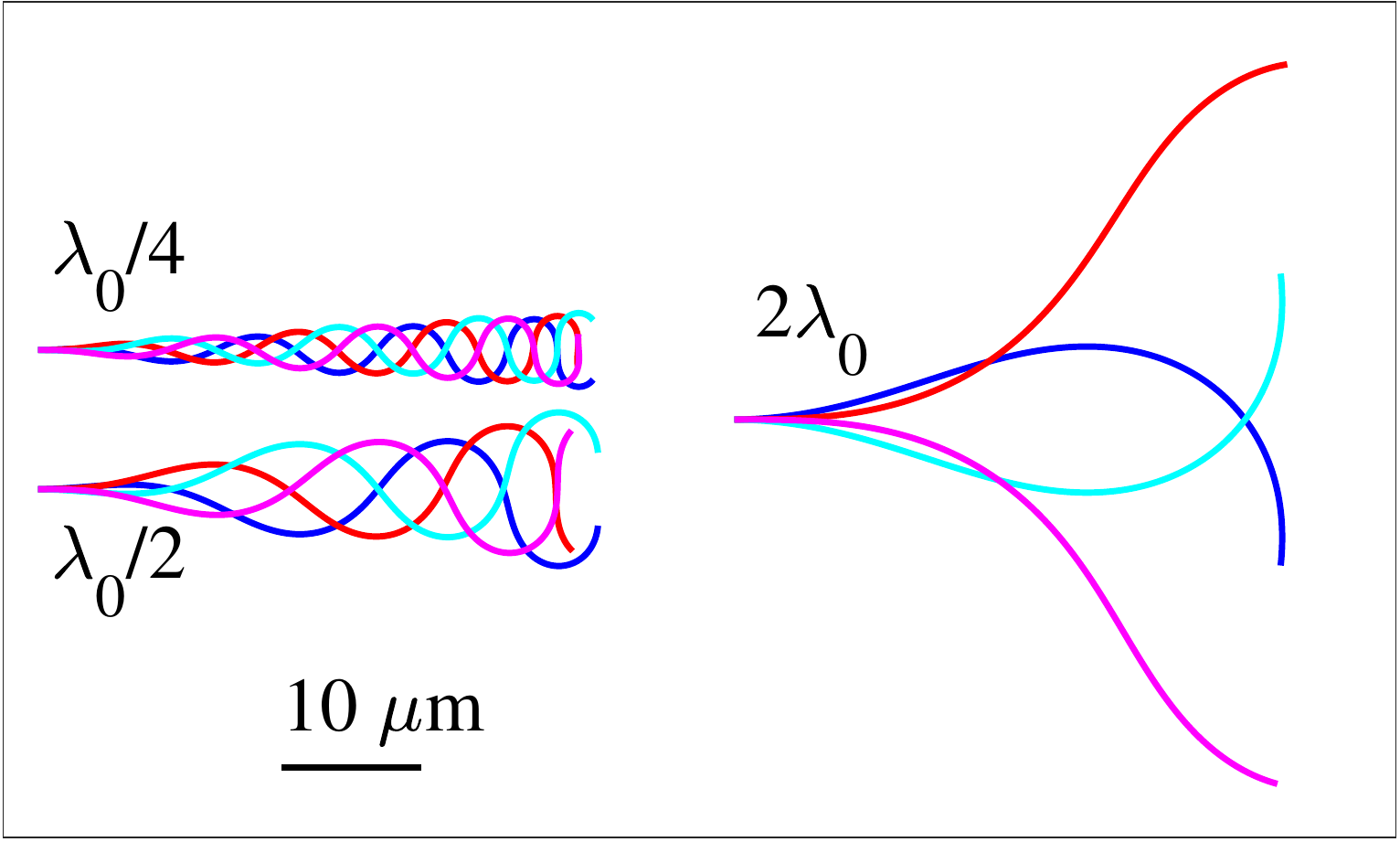}}\\
\end{tabular}
\caption{\label{swimmer_shape81}
{\it (Color  online)}  {\it Left:} Swimmer body shape at six evenly-spaced times within one period of $\pi 10^{-2}$ s. The head is spherical of diameter $L_{head}=$5e-06 m, the flagellum length is 50e-06 m, and the tail beating is given by formula (\ref{beating}). This shape corresponds to the reference wavelength $\lambda_0$. {\it Right:}  Flagellum centerline for four evenly-spaced times within one beating period and three different wavelengths: $\lambda$ = $\lambda_0$/4, $\lambda_0$/2, 2$\lambda_0$.} 
\end{figure}


We denote the mean flagellar curvature as $K_0$, the flagellum frequency as $\nu$, $\lambda$ the wavelength and $A_0$ the amplitude of the wave \cite{Friedrich10}. Following \cite{Zaitsev}, we first use the values: 
$K_0         = 7735.5$ rad/m,
$\nu      = 200$ rad/s,
$\lambda     = \lambda_0= 52.19 \mu$m,
$A_0         = 16828.83$ rad/m, which give a similar flagellum waveshape compared to \cite{Alouges13}. Additionally, we consider the flagellum radius to be $F_r = 0.25\mu$m, and the spherical swimmer head to have diameter of $L_{head} = 5 \mu$m, see Fig. \ref{swimmer_shape81}. We then study the case $K_0=0$ which corresponds to a swimmer following a rectilinear rather than circular trajectory. Furthermore, to investigate the effect of the characteristic wavelength on the swimmer dynamics we perform some simulations with different $\lambda$: $\lambda_0/4$, $\lambda_0/2$, $2\lambda_0$, $\lambda\rightarrow\infty$, see Fig. \ref{swimmer_shape81} (right).

Following \cite{Alouges13} and \cite{Friedrich10}, for an angle $\Psi$ measured along the flagellum arclength $s$ equal to 
\begin{align}
\Psi(s,t)&=K_0 s +2A_0 s \cos\left(\nu t-\frac{2\pi s}{\lambda}\right),\label{Psi}
\end{align}
the flagellum coordinates are 
\begin{align}
{\bf r}(s,t)=\frac{L_{head}}{2}{\bf e}(t) \label{beating}
+\int_0^s{\cos[\Psi(u,t)]{\bf e}(t)+\sin[\Psi(u,t)]{\bf e}^{\perp}(t)}du
\end{align}
where ${\bf e}=[\cos \theta, \sin \theta]$ and ${\bf e}^{\perp}=[-\sin \theta, \cos \theta]$ with $\theta$ the angle between the chosen reference frame and the swimmer reference frame, the latter is shown in Fig. \ref{swimmer_shape81}. In the frame of reference of the swimmer, base vectors are time independent and equal to  ${\bf e}=[1, 0]$ and ${\bf e}^{\perp}=[0, 1]$, hence, the velocities of the flagellum are
\begin{align}
u_x^{BC}(s,t)=-\int_0^s\sin[\Psi(u,t)]\dot{\Psi} du, \label{beating_vel}
\hspace{0.3cm} u_y^{BC}(s,t)=\int_0^s\cos[\Psi(u,t)]\dot{\Psi} du,
\end{align}
with $$\dot{\Psi}=-2A_0\nu s \sin\left(\nu t-\frac{2 \pi s}{\lambda}\right),$$
these values, together with the null-velocity distribution on the swimmer's head, form the known term of the system of equations (\ref{full_system}).

As a remark, we stress that the swimming problem is typically solved in the frame of reference of the swimmer ({\it e.g.} in \cite{Alouges13}) with the origin of the axes on the center of the head as shown in Fig. \ref{swimmer_shape81}. Although arbitrary, this choice is convenient since this is the natural frame to express the velocities of the flagellum. Note, however, that the center of mass of the head is not the center of mass of the entire body, the latter moves as the flagellum itself changes shapes and mostly falls outside the swimmer body. A different choice of the frame of reference leads to different values of forces, torques and velocities, but once the results are recasted in a common frame, {\it e.g.} the laboratory frame, velocities and trajectories coincide. This is a consequence of the fact that the torque for systems with null-force resultants is independent of the location of the frame of reference. In this case the torque resultant is zero too.

When $\lambda\rightarrow\infty$ the expressions for the coordinates simplify and the integrals in $du$ can be easily computed
\begin{equation}
{\bf r}(s,t)=\left[\frac{L_{head}}{2}+\frac{\sin (s \kappa)}{\kappa}\right]{\bf e}+\left[\frac{1}{\kappa}-\frac{\cos (s \kappa) }{\kappa}\right]{\bf e}^{\perp},\nonumber
\end{equation}
where $\kappa=K_0+2A_0\cos(\nu t)$. The boundary conditions result in the complex time-dependent functions
\begin{align}
u_x^{BC}(s,t)&=2A_0\nu\sin(\nu t)\left[\frac{\sin (s \kappa)}{\kappa^2}-\frac{s \cos(s \kappa)}{\kappa}\right],\\
u_y^{BC}(s,t)&=2A_0\nu\sin(\nu t)\left[\frac{1}{\kappa^2}-\frac{\cos (s \kappa)}{\kappa^2}-\frac{s \sin(s \kappa)}{\kappa}\right],\nonumber
\end{align}
where $\cos(s\kappa)$ and $\sin(s\kappa)$ can be expanded in a Taylor series to yield series of respectively even or odd powers of $s\kappa$. In the special case of the rectilinear swimmer, {\it i.e.} $K_0=0$, the expansion simplifies into even, for $u_x^{BC}$, and odd, for $u_y^{BC}$, power series of $2A_0s\cos(\nu t)$ that correspond, in the frequency domain, to spectra  with only even or odd modes different from zero.

After drawing the flagellum center line we use the local Frenet-Serret frame to build the cylindrical surface of radius $F_r$, which we discretize by approximately evenly spaced points. The distance between the points is chosen in such a way that the corresponding surface area approximately equals the surface area $\mathcal{A}$ relative to the points on the head (see end of \ref{Theswimmingproblem}). In conclusion, the regularization parameter for the entire swimmer surface is $\epsilon=0.5\sqrt{\mathcal{A}}$.

\subsection{Finite element method}\label{subsec:FEM}
\begin{figure} 
\centering
\resizebox{4.4cm}{!}{\includegraphics{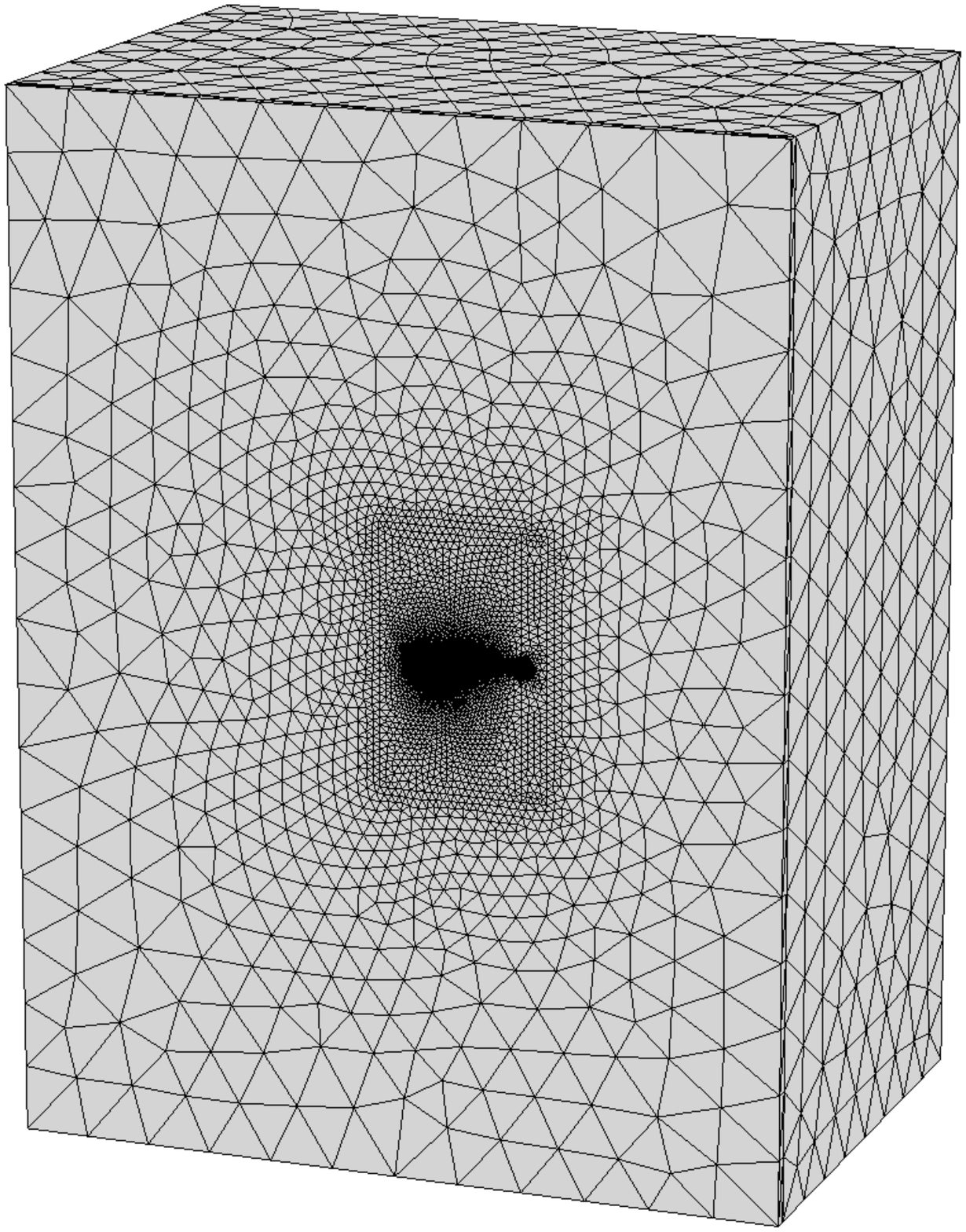}}
\centering
\resizebox{4.4cm}{!}{\includegraphics{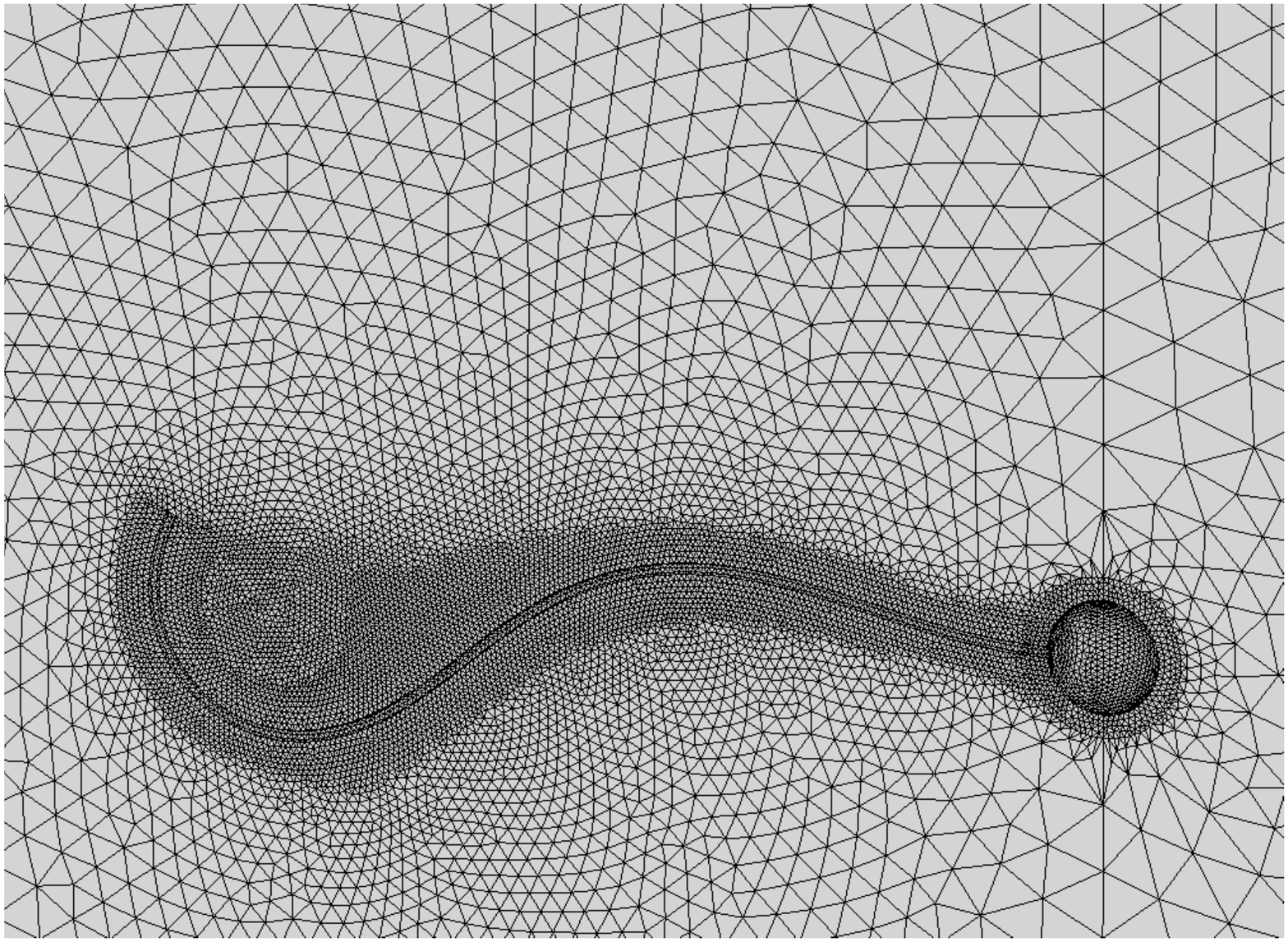}}
\centering
\resizebox{4.4cm}{!}{\includegraphics{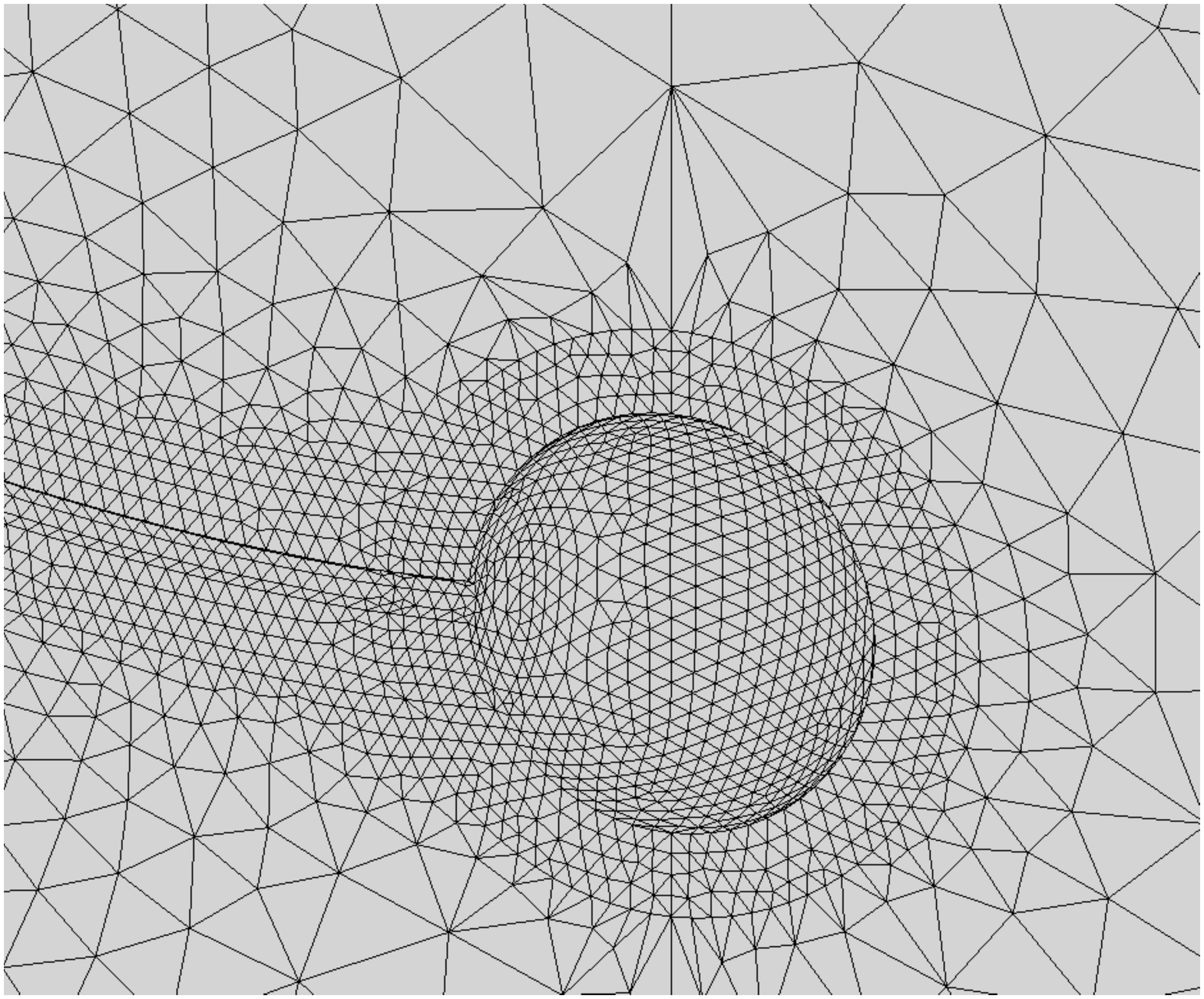}}
\caption{\label{s01}
{{\it Left:} Computational domain for the finite-element solution. {\it Middle:} Zoomed finite-element mesh around the swimmer. {\it Right:} Zoomed finite-element mesh around the swimmer head.}}
\end{figure}

\begin{figure} \centering
\resizebox{10cm}{!}{\includegraphics{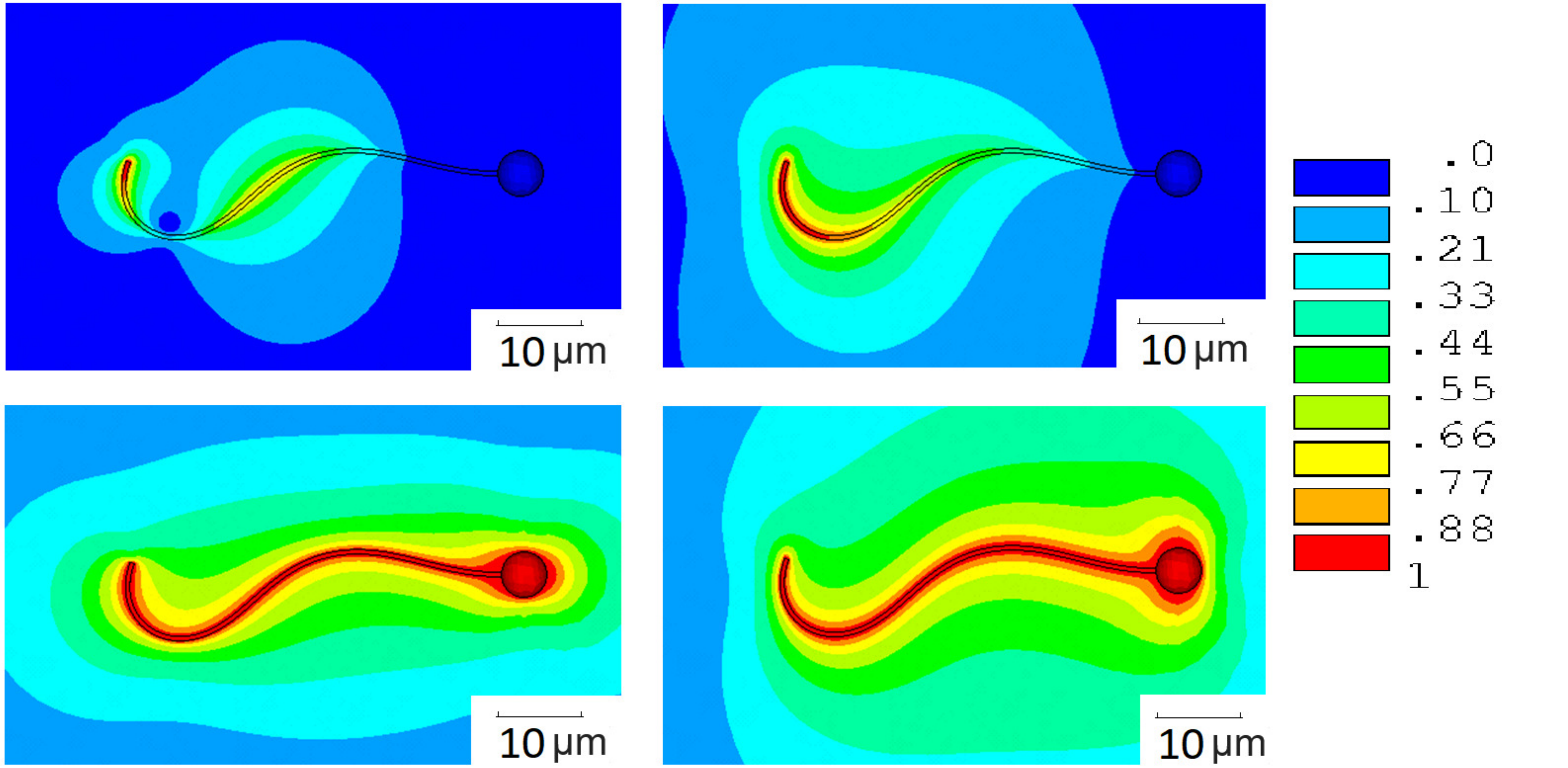}}
\caption{\label{mag_u}
{Absolute velocity distribution in-plane of the swimmer for 4 elementary motions: top left, a flexibly moving flagellum in the reference frame fixed with the swimmer head, top right, the swimmer rotating about the head centre as a rigid body, bottom left, the swimmer is rectilinearly moving in the streamwise direction as a rigid body, bottom right, the swimmer is rectilinearly moving in the transverse direction as a rigid body.}}
\end{figure}

To cross-verify solutions obtained with the regularised Stokeslet method, the system of the governing three-dimensional Stokes equations\\
\begin{equation}
-\nabla p+\mu \nabla^2\bf{u} = 0,
\label{Solid1}  \end{equation}
 \begin{equation}
\nabla \cdot \bf{u} = 0,
\label{Solid2}  \end{equation}
is solved numerically in the reference frame fixed with the centre of the spherical head of the swimmer. The time period of the flagellum motion is discretised into 100 uniform time steps, which amount was found sufficient for accuracy. Each time moment corresponds to a particular configuration of the flagellum wave shape (\ref{Psi})-(\ref{beating}). For each shape of the swimmer, the same open-boundary computational box domain around the swimmer is specified. The box size is large enough to simplify the specification of numerical boundary conditions at the external boundaries. The domain is discretised by tetrahedral grid elements with applying a sufficient refinement near the swimmer boundary to resolve both the head and the flagellum surface. The total number of grid cells in the model is about 440,000 and the grid details are shown in Fig. \ref{s01}. The grid is generated by using the software \MYhref[black]{http://gmsh.info}{"gmsh"}. \\
For each wave form configuration, a converged flow solution is obtained with applying non-slip condition on the swimmer surface with the velocity of the fluid equal to the velocity of the swimmer boundary, which consists of the rigid head and the flexible flagellum parts, and the full slip condition at all external boundaries. The solution obtained is found to be virtually insensitive to any further increase of the computational domain size or a further grid refinement. By integrating the forces on the flagellum surface the drag force components and the torque specified on the right-hand-side of equations (\ref{systemswimming1})-(\ref{systemswimming3}) is calculated. In a similar way, the coefficients for the left-hand-side of the same equations, which correspond to the two elementary rectilinear motions in-plane of the swimmer and the elementary rotation of the swimmer around its head centre as of a rigid body is computed. These amount to 4 boundary value problems, which correspond to the same governing equations (\ref{Solid1})-(\ref{Solid2}), the same computational domain, but different boundary conditions. These 4 problems are solved numerically with a finite-element method for each time moment in accordance with a particular phase of the swimming cycle (Fig. \ref{mag_u}).\\\\
Details of the finite-element method methods for numerical solution are summarised below.  Following the standard approach \cite{Brezzi} the finite-element method with second-order base functions is implemented in the framework of the penalty method, which requires minimisation of the following functional\\
\begin{align}
J(u,v,w)&=\lambda\int\limits_V (\Delta)^2dV
+2\mu\int\limits_V (\epsilon^2_{xx}+\epsilon^2_{yy}+\epsilon^2_{zz}+\frac12\epsilon^2_{xy}+\frac12\epsilon^2_{xz}+\frac12\epsilon^2_{yz})dV\nonumber \\
&-\int\limits_V (f_xu+f_yv+f_zw)dV \nonumber
\end{align}
with penalty parameter $\lambda$ ,  where $\epsilon_{xx}, \epsilon_{yy}, \epsilon_{zz}, \epsilon_{xy}, \epsilon_{xz}, \epsilon_{yz}$
    are components of the strain rate tensor,  
\begin{align}
(\epsilon_{xx}, \epsilon_{yy}, \epsilon_{zz}, \epsilon_{xy}, \epsilon_{xz}, \epsilon_{yz})&=\left[\frac{{\partial  u}}{{\partial x}} , \frac{{\partial  v}}{{\partial y}}, \frac{{\partial  w}}{{\partial z}},\right. 
\left. \frac12\left(\frac{{\partial  u}}{{\partial y}}+\frac{{\partial  v}}{{\partial x}}\right), 
\frac12\left(\frac{{\partial  u}}{{\partial z}}+\frac{{\partial  w}}{{\partial x}}\right), 
\frac12\left(\frac{{\partial  v}}{{\partial z}}+\frac{{\partial w}}{{\partial y}}\right)\right];\nonumber \\
\Delta&=\frac{{\partial  u}}{{\partial x}}+\frac{{\partial  v}}{{\partial y}}+\frac{{\partial  w}}{{\partial z}}; \nonumber
\end{align}
and  $f_x, f_y, f_z$  are internal forces. This results in a sparse system of linear algebraic equations that is solved using a direct method based on LU decomposition. The \MYhref[black]{https://software.intel.com/en-us/mkl}{Intel Math Kernel Library} solver is used for solution of the linear system of equations.

\section{Numerical results and data analysis} \label{sec:numerics}

\begin{figure} \centering
\resizebox{5.5cm}{!}{\includegraphics{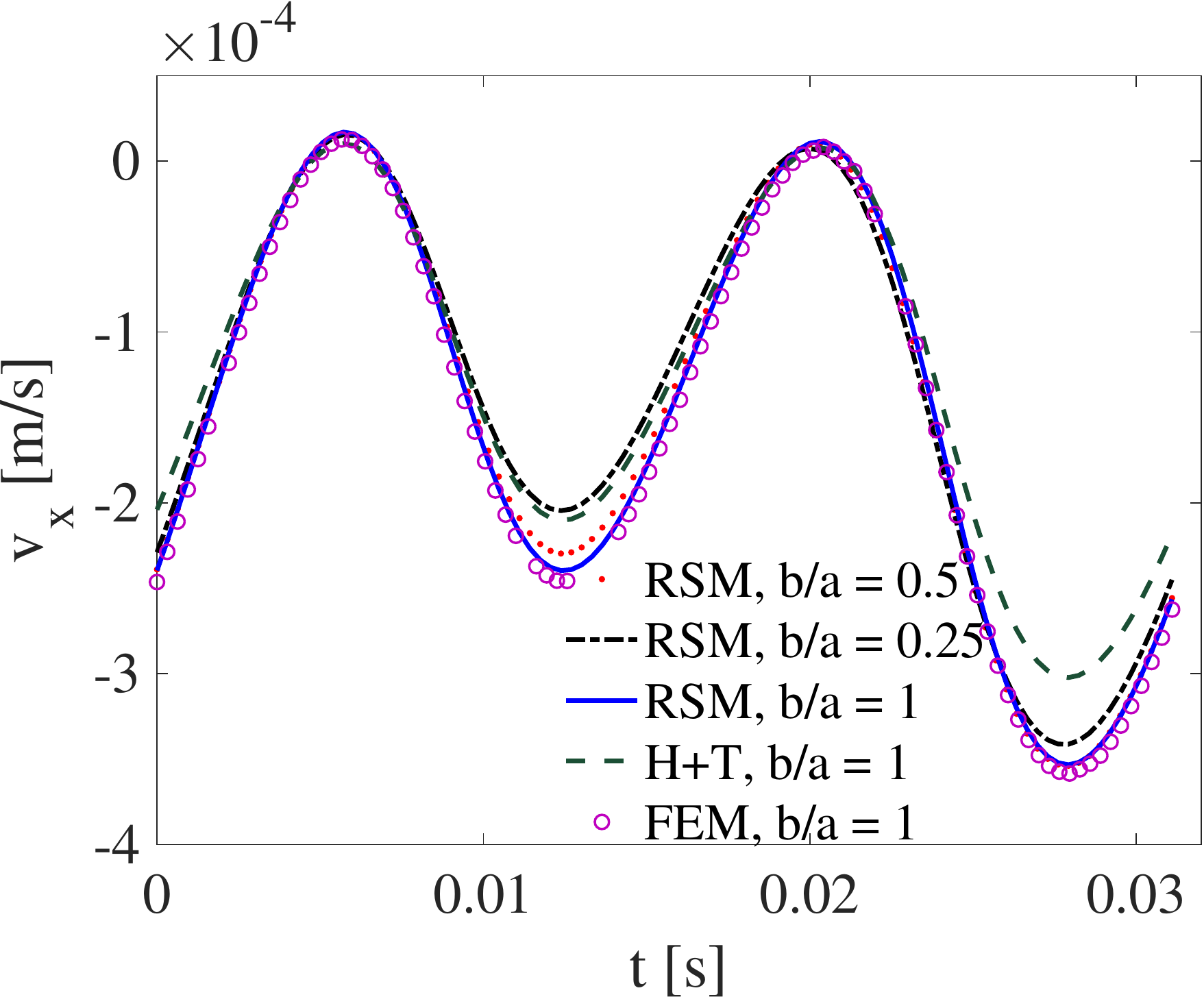}}
\resizebox{5.5cm}{!}{\includegraphics{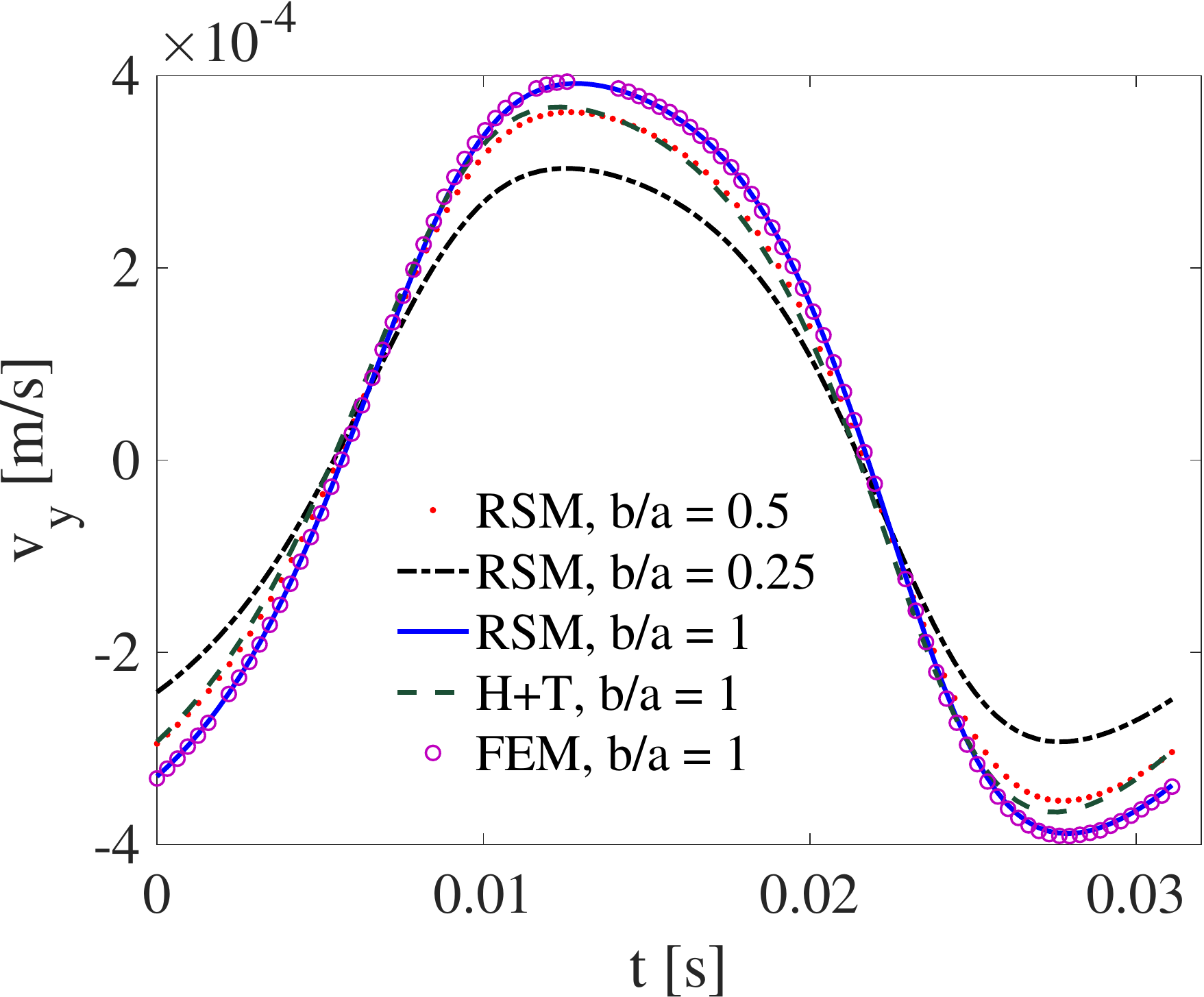}}
\resizebox{5.5cm}{!}{\includegraphics{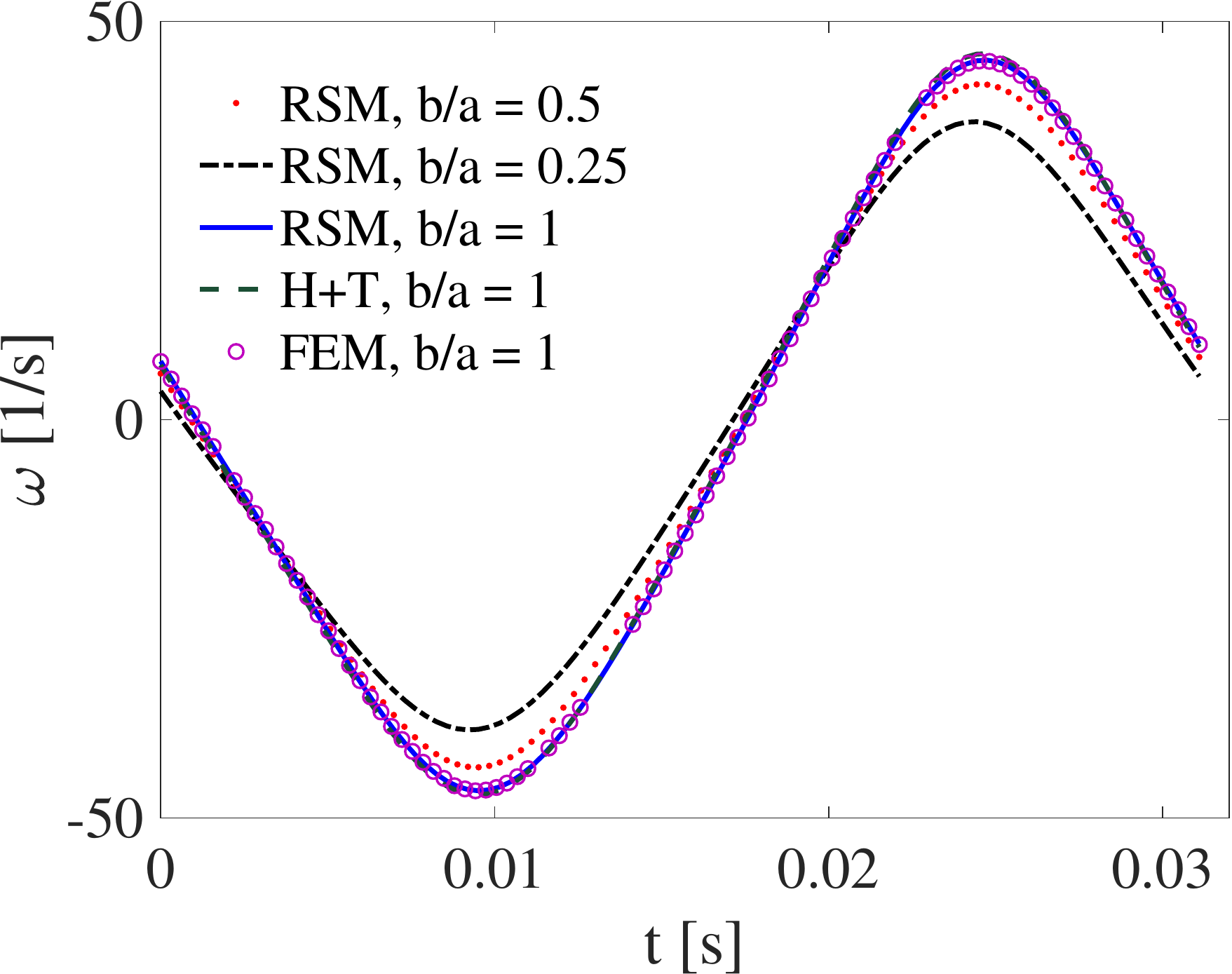}}
\caption{\label{vx}
{\it (Color  online)} $v_x$, $v_y$ components of the velocity in the swimmer frame of reference (left and center panel) and angular velocity (right panel) for a cell with a spherical head  (solid-blue, circle magenta and dashed-green) or an ellipsoidal head with aspect ratio 0.5 (dotted-red) and 0.25 (dashed-dot black). The circle-magenta points have been computed by a FEM code for comparison, the dashed-green curve were computed by using the simplified approach discussed in section \ref{simplified}. The frame of reference for these calculations is located on the head centroid.}
\end{figure}
\begin{figure} \centering
\resizebox{7.5cm}{!}{\includegraphics{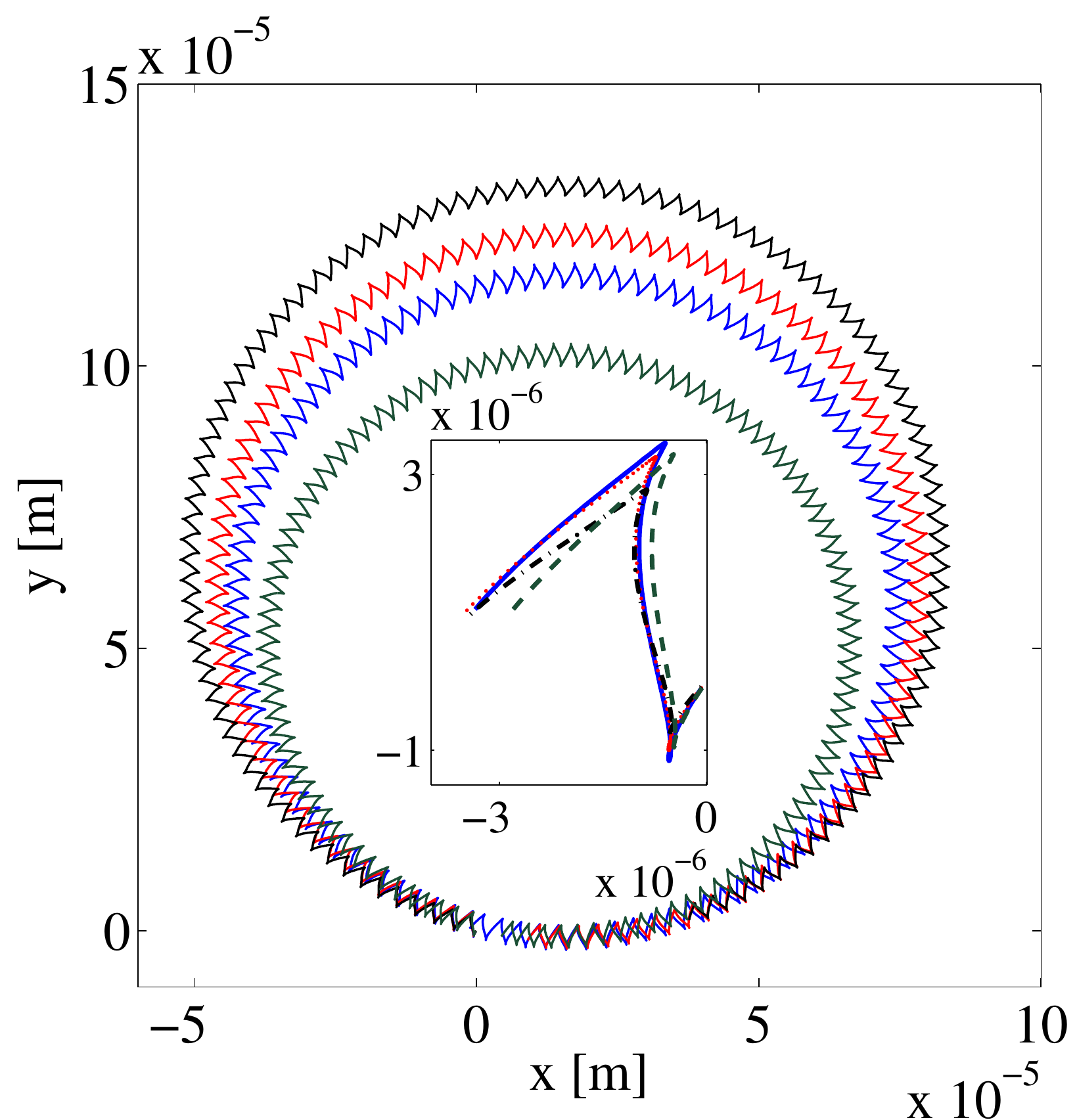}}
\caption{\label{trajectory}
Swimming trajectory on the $x$-$y$ plane for 107 beating periods and one period (inset) for a swimmer with a spherical head (solid blue) and a swimmer with an ellipsoidal head with aspect ratio equal to 0.5 (dotted red) and 0.25 (dashed-dotted black). In dashed-green, for comparison, the trajectory of a swimmer with a spherical head computed by approximating the resistive matrix coefficients as the sum of the head and flagellum separately. Note the smaller radius of the trajectory corresponding to larger coefficients $C_x^{v_x}$, $C_y^{v_y}$ (see Table \ref{Headon-offTable}). These are the trajectories for the head centroid, the axes units are in meters.}
\end{figure}

In Fig. \ref{vx} we plot the $x$ and $y$ component of the velocity and the $z$ component of the angular velocity in the frame of reference of the swimmer for different numerical methods and head shapes. Different points on the swimmer body draw different trajectories on the $x$-$y$ plane, in Fig. \ref{trajectory} we display the trajectories of the head centroid. 

An inspection of the frequency spectra of $v_x$, $v_y$ and $\omega$ reveals that the signal can be reconstructed within a 0.6\% error by retaining the first 5 terms of the Fourier series expansion: 
$\sum_{k=0}^{4}a_k \cos( k \nu t+\phi_k)$. The error is computed as: $\max|v-v_{reconstructed}|/\max|v|$, where $v_{reconstructed}$ is the reconstructed signal from the truncated Fourier series and $v$ the original signal. When the parameter $K_0$ is set to zero (the rectilinear swimmer) the curves $v_y$ and $\omega$ have zero mean and are described within a 0.6\% error by the first and third mode; keeping only the mode $k=1$ guarantees an 8\%  error on $v_y$ and a 6\% error on $\omega$. Differently, $v_x$ is described within a 0.3\% error by an expansion in the even modes $k=0,2,4$.  
We have additionally verified for the rectilinear swimmer that the temporal variation of the angular frequency of the swimmer, $\omega$, obtained numerically can be reasonably well approximated (within 6\%) by a single harmonic function of the beating frequency, $\nu$, regardless of the numerical discretisation applied (e.g. 100, 200, and 400 points per the flagellum length).

We stress that since the system is linear no mechanism is in place to allow for the creation of non-zero modes from the forcing/boundary condition, in fact, the non-zero modes detected in $v_x$, $v_y$ and $\omega$ reflect the complex spectrum of $u_x^{BC}$ and $u_y^{BC}$ (eq. \ref{beating_vel}), which produces a coupling of the $x$- and $y$-coordinates of the local reference system of the flagellum. The fact that for the rectilinear swimmer only the even modes are excited in $v_x$ and the odd ones in $v_y$ matches the  $u_x^{BC}$ and $u_y^{BC}$ spectra for the special case $K_0=0$ as discussed at the end of section \ref{subsec:parameters}. Similar considerations hold when comparing the spectra of $F_x^B$, $F_y^B$ $T_z^B$, that is the known term of the resistive matrix system and the unknowns $v_x$, $v_y$, $\omega$.

As a preliminary validation, we compare the results obtained by means of the Regularized Stokeslet Method with those attained through the Finite Element Method \cite{Zaitsev}: see the difference in the resistive matrix coefficients (Table \ref{Reg-FEMTable}), and compare the blue solid line and the magenta circles in Fig. \ref{vx}. The difference between the coefficients is within few percents, while the differences between the velocities, quantified in Table \ref{RFT-table}, are barely distinguishable and of second order when compared with the effects of the head-shape or the errors introduced by the simplified approaches (dashed-green curve). 
This consistency guarantees the accuracy of our results and the reliability of  both the numerical schemes. 

Compared to the RSM, the FEM calculation is much more expensive since it solves the governing equations discretised in the entire flow domain and not just on the swimmer's surface.
For example, the RSM calculations performed here took several minutes per case on a single processor. For the FEM calculation, the same required about 35 hours with running two OpenMP threads in parallel. The amount of computer memory in each case was more comparable: 15Gb for the RSM method and 27 Gb for the FEM solution per case.

We have verified that for the calculations presented in this paper the motility matrix coefficients are symmetric within numerical precision as dictated by the reciprocal theorem.

As a further remark, note that the curvature parameter $K_0>0$ of the swimmer wave form in (\ref{Psi}) corresponds to a circular trajectory in the absolute 
frame of reference as seen in Fig. \ref{trajectory}. Accordingly, this should give rise to apparent accelerations in the swimmer's frame, which are not accounted for in the Stokes model. 
To justify the neglect of these accelerations, we want to evaluate the order of magnitude of these terms first. The difference between the accelerations in the non-inertial frame ${\bf a}_r$ and those in the inertial frame ${\bf a}_f$ are:
\begin{equation}
{\bf a}_r-{\bf a}_f=-{\bm \omega} \times ({\bm \omega} \times {\bf r})-2{\bm \omega}\times{\bf v}-\dot{{\bm \omega}}\times{\bf r}
\end{equation}
 
where the first term on the RHS represents the centrifugal acceleration, the second the Coriolis acceleration and the third the Euler acceleration. The force associated to ${\bf a}_r$ in our calculations is at most of the order $\mathcal{O}$(1e-15), that is three orders of magnitude smaller than the smallest coefficients in the motility matrix, thus negligible as initially hypothesized. Still, it can be argued that even a small unbalanced force can build up into a non-negligible effect for the swimmer trajectory over a time period  long enough compared to the swimmer cycle. Therefore, to confirm that the effect of the non-inertial forces on the trajectory of the swimmer is small, we compared the swimmer's trajectories with and without taking the apparent accelerations into account in accordance with the ``instantaneous" coordinate and velocity of the swimmer calculated numerically. Over a few circular trajectory periods, the swimmer trajectories with and  without taking the apparent accelerations into account virtually coincided, which finally justifies the neglect of these terms.

\begin{table}[!h]
\caption{Difference between the resistive matrix coefficients and known terms of system (\ref{systemswimming1})-(\ref{systemswimming3}) computed with the Regularized Stokeslet Method (RSM) and the Finite Element Method (FEM) for a spherical head swimmer. The difference is computed as $\dfrac{\sum(C_{RSM}-C_{FEM})}{\sum(C_{RSM}+C_{FEM})/2}*100$, where the sum is performed over the time discretization and $C$ refers to any of the coefficients considered. The second row in this table indicates the order of magnitude of the corresponding coefficient.}
\begin{center}
\tabcolsep=0.11cm
\begin{tabular}{ccccccccccccccc}
\hline
$C_x^{v_x}$ & $C_x^{v_y}$ & $C_x^{\omega}$ & $C_y^{v_x}$ & $C_y^{v_y}$ & $C_y^{\omega}$ & $m_z^{v_x}$ & $m_z^{v_y}$ & $m_z^{\omega}$ & $F_x^B$ & $F_y^B$ & $T_z^B$\\
\footnotesize{$\sim$1e-4}
&\footnotesize{$\sim$1e-6}
&\footnotesize{$\sim$1e-10}
&\footnotesize{$\sim$1e-6}
& \footnotesize{$\sim$1e-4}
& \footnotesize{$\sim$1e-9}
&\footnotesize{$\sim$1e-10}
&\footnotesize{$\sim$1e-9}
&\footnotesize{$\sim$1e-14}
&\footnotesize{$\sim$1e-8}
& \footnotesize{$\sim$1e-8}
&\footnotesize{$\sim$1e-12}\\
\hline
 2.92 \%            & -5.5\%          & 1.55\%                & -5.5\%        & 1.49\%        & 0.9\% &   1.55\%               &   0.9\%        & 3.69\%     & -1.76\% & -3.58\% & 1.06\% \\
\hline
\end{tabular}
\end{center}
\label{Reg-FEMTable}
\end{table}%

\begin{table}[!h]
\caption{Values of the time averaged and rescaled root mean square error for the swimming velocities computed with the RFT-L, the RFT-GH, the head+tail (H+T) and the FEM model. The results are for the rectilinear ($K_0=0$ rad/s) and curved ($K_0=7735.5$ rad/s) swimmer with $\lambda=\lambda_0$.}
\label{RFT-table}
\begin{center}
\begin{tabular}{llccc}
\hline
& rad/s&$v_x$ RMS Error & $v_y$ RMS Error & $\omega$ RMS Error\\
\hline
RFT-GH, &$K_0=0$ & 0.1577 & 0.1103 & 0.06478 \\
RFT-L, &$K_0=0$ & 0.08814 & 0.06211 & 0.04796 \\
RFT-GH, &$K_0=7735.5$ & 0.1337 & 0.1102 & 0.06492 \\
RFT-L, &$K_0=7735.5$ & 0.07523 & 0.06196 & 0.04787 \\
H+T, &$K_0=7735.5$ & 0.06963 & 0.03545 & 0.000632 \\
FEM, &$K_0=7735.5$ & 0.01448 & 0.00334 & 0.001277 \\
\hline
\end{tabular}
\end{center}
\end{table}%

\begin{table}[!h]
\caption{Values of the longitudinal friction coefficients for a unitary viscosity (the values from \cite{Friedrich10} and \cite{Alouges13} are divided by 0.7 mPa s) and ratio between the normal and longitudinal friction coefficients for the head and flagellum. Following \cite{Friedrich10} and \cite{Alouges13} the head friction coefficient is computed for an aspect ratio $b/a$=0.5 and $L_{head}$=10$\mu$m.}
\begin{center}
\tabcolsep=0.11cm
\begin{tabular}{lcccc}
\hline
&$C_x^{v_x,h}$ & $C_y^{v_y,h}$/$C_x^{v_x,h}$  & $K_T$ & $K_N$/$K_T$\\
& N s/m & - & N s/m$^2$ & -\\
\hline
Stokes and RFT-L    &  5.6734e-05                    &  1.14532  &  1.7326 & 1.76  \\
Stokes and RFT-GH & 5.6734e-05                     &  1.14532  & 1.1353  &  1.69      \\
Perrin's formula and $K_N$ and $K_T$ as in \cite{Friedrich10} & 5.7571e-05 &   1.14392 &  0.9857$\pm$0.8857       &  1.81$\pm$0.07   \\
Perrin's formula and $K_N$ and $K_T$ as in \cite{Alouges13} & 5.7571e-05 &   1.14392 &  0.5429       &  1.89   \\
\hline
\end{tabular}
\end{center}
\label{RFTratio}
\end{table}%

\subsection{Resistive Force Theory}\label{RFT}
According to the Resistive Force Theory (RFT) the viscous forces applied to the flagellum centerline depend on the flagellum velocities through
\begin{equation}
{\bf f}=[K_N\mathcal{I}-(K_T-K_N) {\bf t}{\bf t}^T]\cdot {\bf u}^{BC},
\label{RFTeq}
\end{equation}
where $K_T$ and $K_N$ are the tangential and normal friction coefficients, ${\bf t}=(\cos \Psi, \sin \Psi)$ the tangent to the flagellum centerline and $\mathcal{I}$ the identity matrix. Equation (\ref{RFTeq}) can alternatively be expressed as \cite{Zaitsev}
\begin{equation}
{\bf f} = \mathcal{R}\mathcal{K}\mathcal{R}^{-1}{\bf u}^{BC}
\label{MihailRFT}
\end{equation}
where $\mathcal{R}$ is the rotation matrix 
\begin{equation}
\left[
\begin{array}{cc}
\cos\Psi & -\sin\Psi  \\
\sin\Psi & \cos\Psi  \\
\end{array} 
\right]\label{Rot}
\end{equation}
and $\mathcal{K}$ is a diagonal matrix with $K_T$ and $K_N$ on the diagonal. 
We proceed by ({\it i}) substituting the expressions for $\Psi$ and ${\bf u}^{BC}$ provided in Section \ref{subsec:parameters} in (\ref{RFTeq}) or (\ref{MihailRFT}), ({\it ii}) computing the coefficients of the propulsion matrix and ({\it iii}) solving the linear system (\ref{systemswimming1})-(\ref{systemswimming3}). In point ({\it ii}) the friction coefficients are computed as the sum of the head and tail contribution as expressed by (\ref{tail+head}). The head contribution is known analytically for a spherical or ellipsoidal head, while the tail friction coefficients are computed by integrating numerically (\ref{RFTeq}) or (\ref{MihailRFT}) and the $z$ component of the torque ${\bf x} \times {\bf f}$ for the entire flagellum length after replacing ${\bf u}^{BC}$ by, in turn, a unitary forward and transversal velocity and an unitary angular velocity.

\begin{figure} \centering
\resizebox{8cm}{!}{\includegraphics{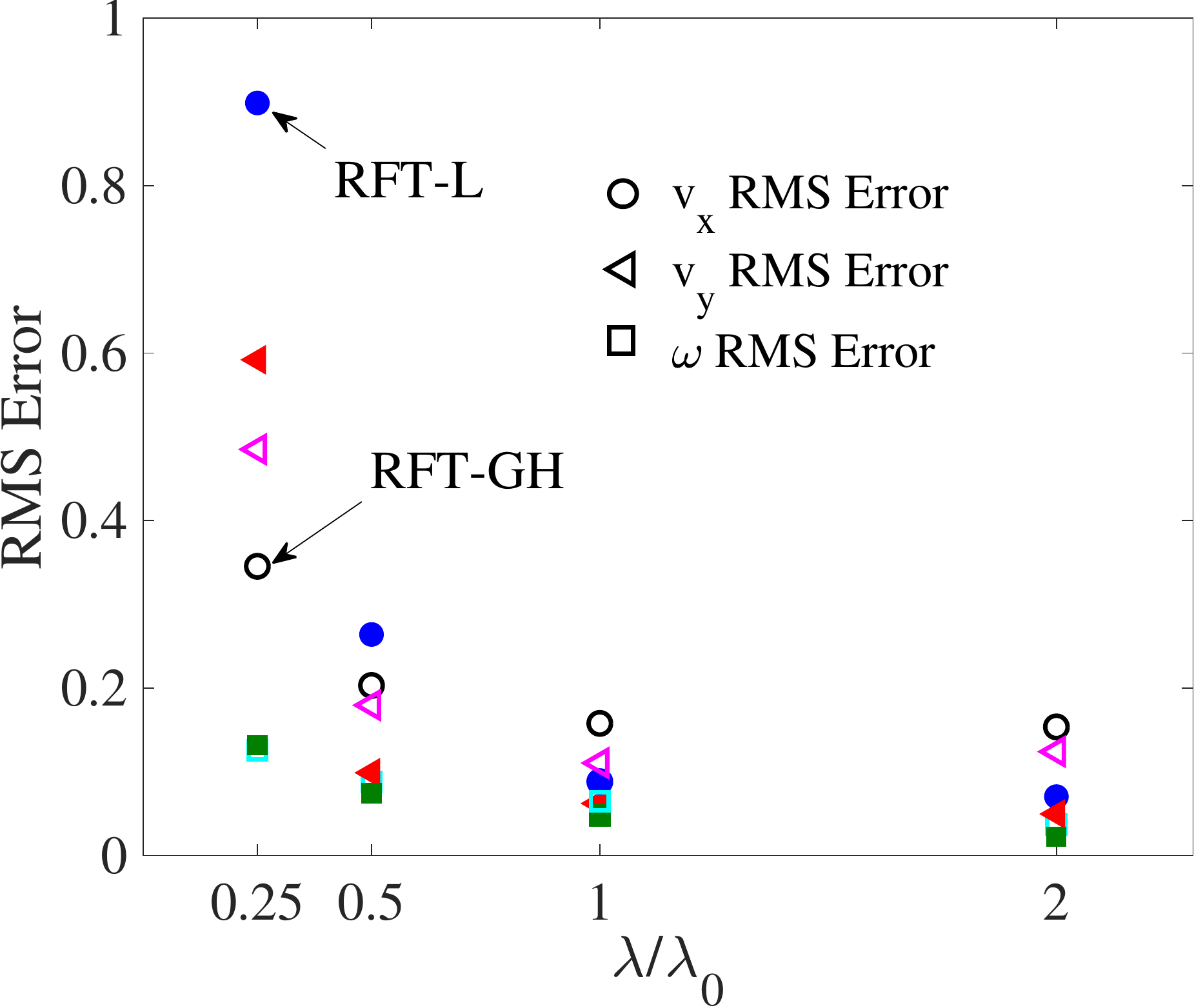}}
\caption{\label{vx_approx}
{\it (Color  online)} Root mean square error averaged over one period for the swimming velocities ($v_x$-circles, $v_y$-triangles) and the angular velocity ($\omega$-squares) for the RFT Lighthill (RFT-L, filled symbols) and the RFT Gray and Hancock (RFT-GH, empty symbols) model. The error is plotted as a function of the wavenumber $\lambda$ normalized by the reference wavenumber $\lambda_0$ [see formula (\ref{Psi})]. The results are for the rectilinear swimmer. In Table \ref{RFT-table} we report the numerical value of the errors for $\lambda=\lambda_0$ and some extra cases.}  
\end{figure}

Following \cite{Rodenborn13} we contrast the results for two possible choices of the friction coefficients: those derived by Lighthill (RFT-L) \cite{Lighthill75} and those suggested by Gray and Hancock (RFT-GH) \cite{Gray55} 
\begin{align}
K_{T,L}    &= \frac{2\pi\mu}{\ln\left(\frac{0.18\lambda}{F_r}\right)},             & K_{N,L}    &=\frac{4\pi\mu}{\ln\left(\frac{0.18\lambda}{F_r}\right)+\frac{1}{2}}, \\          
K_{T,GH}&= \frac{2\pi\mu}{\ln\left(\frac{2\lambda}{F_r}\right)-\frac{1}{2}}, & K_{N,GH} &= \frac{4\pi\mu}{\ln\left(\frac{2\lambda}{F_r}\right)+\frac{1}{2}}.
\end{align}
We also compare the results obtained with the RFT with those obtained with the Regularized Stokeslet Method. 
In Fig. \ref{vx_approx} and Table \ref{RFT-table} we quantify the error of the former as $$\mbox{RMS Error} = \frac{\sqrt{<(u_{RFT}-u_{Stk})^2>}}{(\max u_{Stk}-\min u_{Stk})},$$ where `$<>$' average over one beating period, $u_{Stk}$ refers to the value computed with the regularized stokeslet model, and the denominator rescales the root mean square error by the range of variability of the quantity under consideration, being it $v_x$, $v_y$ or $\omega$. For the reference case of the swimmer with $\lambda=\lambda_0$ (solid blue line in Fig. \ref{vx}) we find that Lighthill coefficients outperform Gray and Hancock's both for the curved and rectilinear swimmer, see Fig. \ref{vx_approx} and Table \ref{RFT-table}. However, the model performances depend on the geometry, as clearly seen in Fig. \ref{vx_approx}. For $\lambda<\lambda_0$ we find that the RFT-GH model gives better answers than RFT-L and for $\lambda=\lambda_0/4$ both RFT models are deemed unreliable. We have also checked that the simplified model referred to as the head+tail model (H+T), which will be discussed in the next section, performs better than the RFT (see Table \ref{RFT-table}). 

Note that, as stressed in \cite{Friedrich10}, what really matters to reproduce the kinematics of the motion correctly is the ratio between the tangential and the normal friction coefficients, $K_N/K_T$. The choice of the friction coefficients for the head is also only important in relative terms: the absolute values of $C_x^{v_x}$ and $C_y^{v_y}$ are irrelevant as far as they maintain the right proportion with one another and the flagellum coefficients, this is because the kinematics results from a force balance. The choice of using Perrin's formulae for the brownian motion of an ellipsoid provides coefficients whose ratio is comparable to the ratio for an ellipsoid in a Stokes flow as given by (\ref{cxvxhe})-(\ref{cyvyhe}) and whose absolute values (at least for the $b/a$=0.5 aspect ratio) are not very dissimilar, see Table \ref{RFTratio}. However, given the size and speeds involved, resorting to the Stokes law seems more physically based. Table \ref{RFTratio} reveals that only the RFT-L model has both $K_T$ and $K_N/K_T$ within the range indicated in \cite{Friedrich10}.
While in \cite{Friedrich10} it is found that the RFT reproduces the trajectories reliably, in \cite{Rodenborn13} it is reported that the drag values computed with RFT are inaccurate. This may not necessarily be in contrast since for the last calculations it is the absolute value of the coefficients that matters. However, it is most likely the case that while the customarily chosen RFT coefficients well fit the data for the swimming of a spermatozoon with reference parameters, they do not match the results for different geometries, {\it e.g.} larger or smaller $\lambda$s or helical flagella as in \cite{Rodenborn13}.

\begin{figure} \centering
\resizebox{6.5cm}{!}{\includegraphics{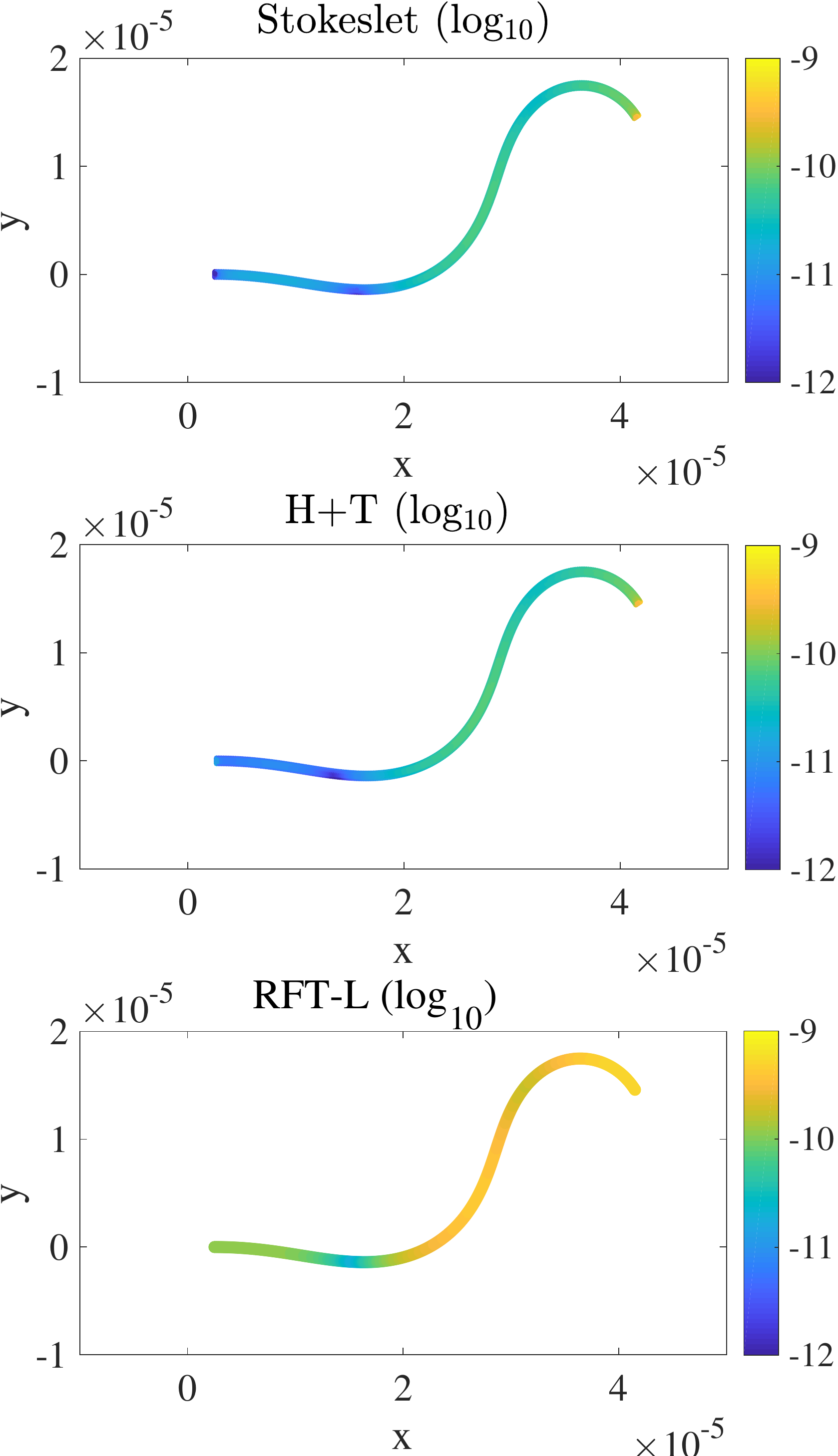}}
\caption{\label{f_tail}
Logarithm with base 10 of the modulus of the forces on the surface of the flagellum for the full-body regularized stokeslet model (top), the head+tail model (middle) and the Resistive Force Theory with Lighthill's coefficients (bottom). For the first two models the forces are distributed on the cylindrical surface of the flagellum whose radius is $F_r=0.25\mu$m, while for the RFT forces are distributed on the flagellum centerline. Figures are in logarithmic scale to highlight the force distribution on the entire extension of the tail, note in fact that the values at the tip of the tail are the largest and a linear scale would only stress this result.}
\end{figure}

\subsection{Inaccuracy of treating the head and flagellum separately and further comparisons with the RFT solutions}\label{simplified}
Although being accurate compared to the RFT, our results indicate that the simplified model consisting in evaluating the friction coefficients as the sum of the head and the tail contribution calculated independently, leads to notable errors. This approximation is convenient for first estimates since it reduces the computational cost when analytical solutions are available, {\it e.g.} for spherical or ellipsoidal heads, 
but neglects the interaction between the tail and the head. The system (\ref{full_system}), in short, ${\bf u^{BC}+v_j+\omega\times{\bf x}\cdot e_j}=\bf u^{tot}=\bf{\mathcal{J}}{\bf f}$, is inverted into ${\bf f}=\bf{\mathcal{J}^{-1}}{\bf u^{tot}}$, in this form each component of the vector of local forces $\bf{f}$, being it located on the head or on the tail, can be split up into two contributions: one due to the points located on the head and one due to the points located on the tail. With superscript $h$ and $t$ denoting the head and tail respectively and greek letters indicating the points on the surface of the swimmer we have $$f_{\alpha^h}=\mathcal{J}^{-1}_{\alpha^h\beta^h}u^{tot}_{\beta^h}+\mathcal{J}^{-1}_{\alpha^h\beta^t}u^{tot}_{\beta^t}$$ and $$f_{\alpha^t}=\mathcal{J}^{-1}_{\alpha^t\beta^h}u^{tot}_{\beta^h}+\mathcal{J}^{-1}_{\alpha^t\beta^t}u^{tot}_{\beta^t},$$ the approximate approach neglects the head-tail interaction terms: $\mathcal{J}^{-1}_{\alpha^h\beta^t}u^{tot}_{\beta^t}$ and $\mathcal{J}^{-1}_{\alpha^t\beta^h}u^{tot}_{\beta^h}$. 

For the case of a spherical head, we have verified that the approximate method  overestimates the leading coefficients of eq. (\ref{systemswimming1})-(\ref{systemswimming2}), $C_x^{v_x}$ and $C_y^{v_y}$, by approximately 23.5\% and 17\%  and incorrectly capture 
many others, see Table \ref{Headon-offTable}. The difference between the resistive matrix coefficients calculated with the two methods is expressed by 
\begin{equation}
\frac{\int_{0}^{T} (C_{\text{head-off}}-C_{\text{head-on}}) \text{dt}}{\int_{0}^{T} C_{\text{head-on}} \: \text{dt}}*100, \label{errorCoeff}
\end{equation}
where $C$ refers to any coefficient and known term of eq. (\ref{systemswimming1})-(\ref{systemswimming3}). A positive value signifies that on average the simplified approach overestimates the coefficient. Remarkably, the overestimate of  $C_x^{v_x}$ and $C_y^{v_y}$ leads to swimming trajectories with different radii as shown in Fig. \ref{trajectory} (compare the solid blue and dashed-green curve).

The simplified approach results in a different distribution of the forces on the surface of the flagellum as shown in Fig. \ref{f_tail}, note that the values differ in particular on the left-end where the flagellum is attached to the head. In the full-body model the values are zero because of the presence of the boundary, while in the simplified model the values are relatively large since this is a free end, however they are not as high as on the tip of the tail given the lower velocities. Note again that even if inaccurate, the head+tail model clearly outperforms the Resistive Force Theory (bottom panel of Fig. \ref{f_tail}) for which the force distribution only qualitatively resembles the first two models. 

\begin{figure*} \centering
\resizebox{4.4cm}{!}{\includegraphics{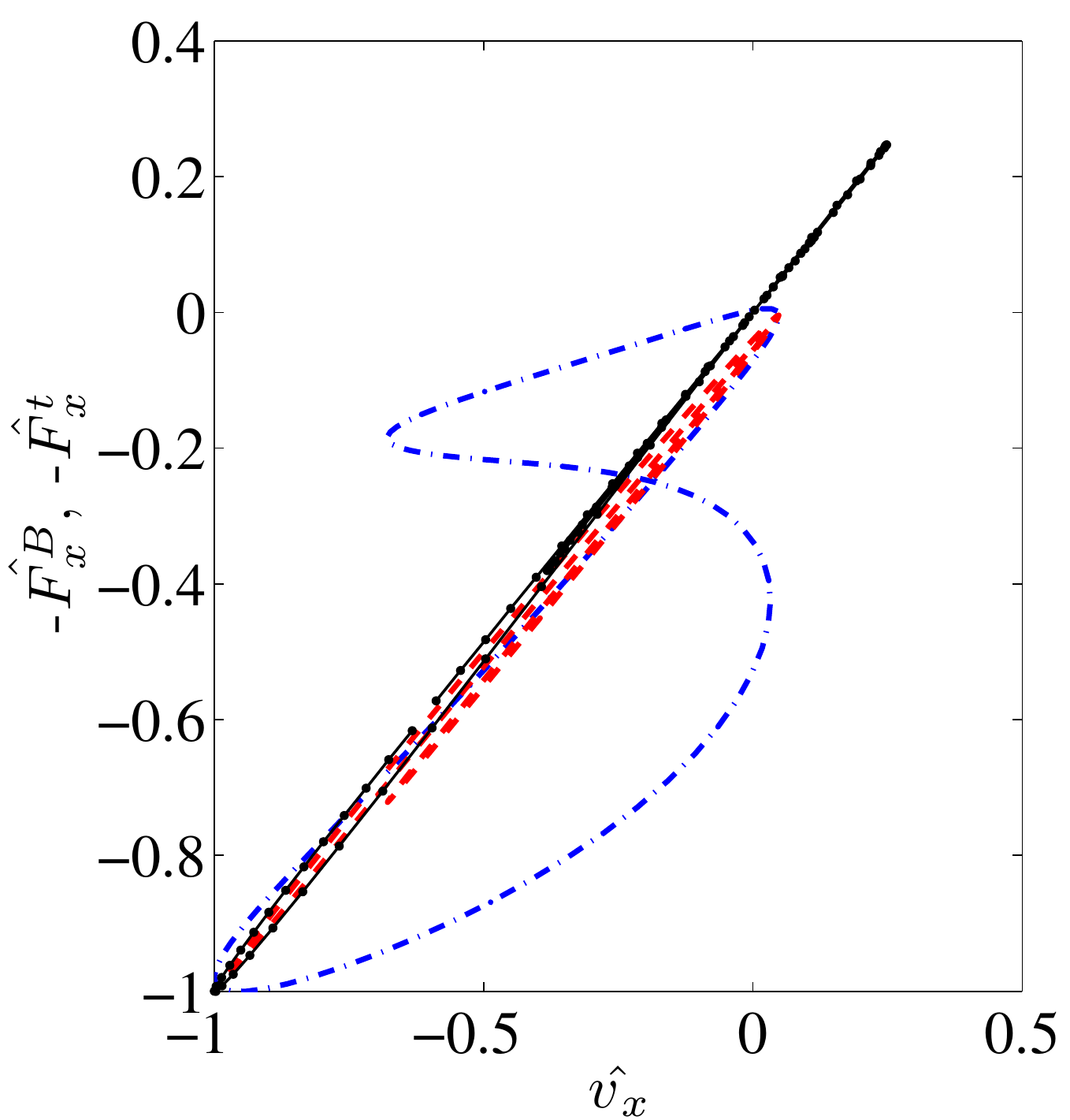}}
\resizebox{4.4cm}{!}{\includegraphics{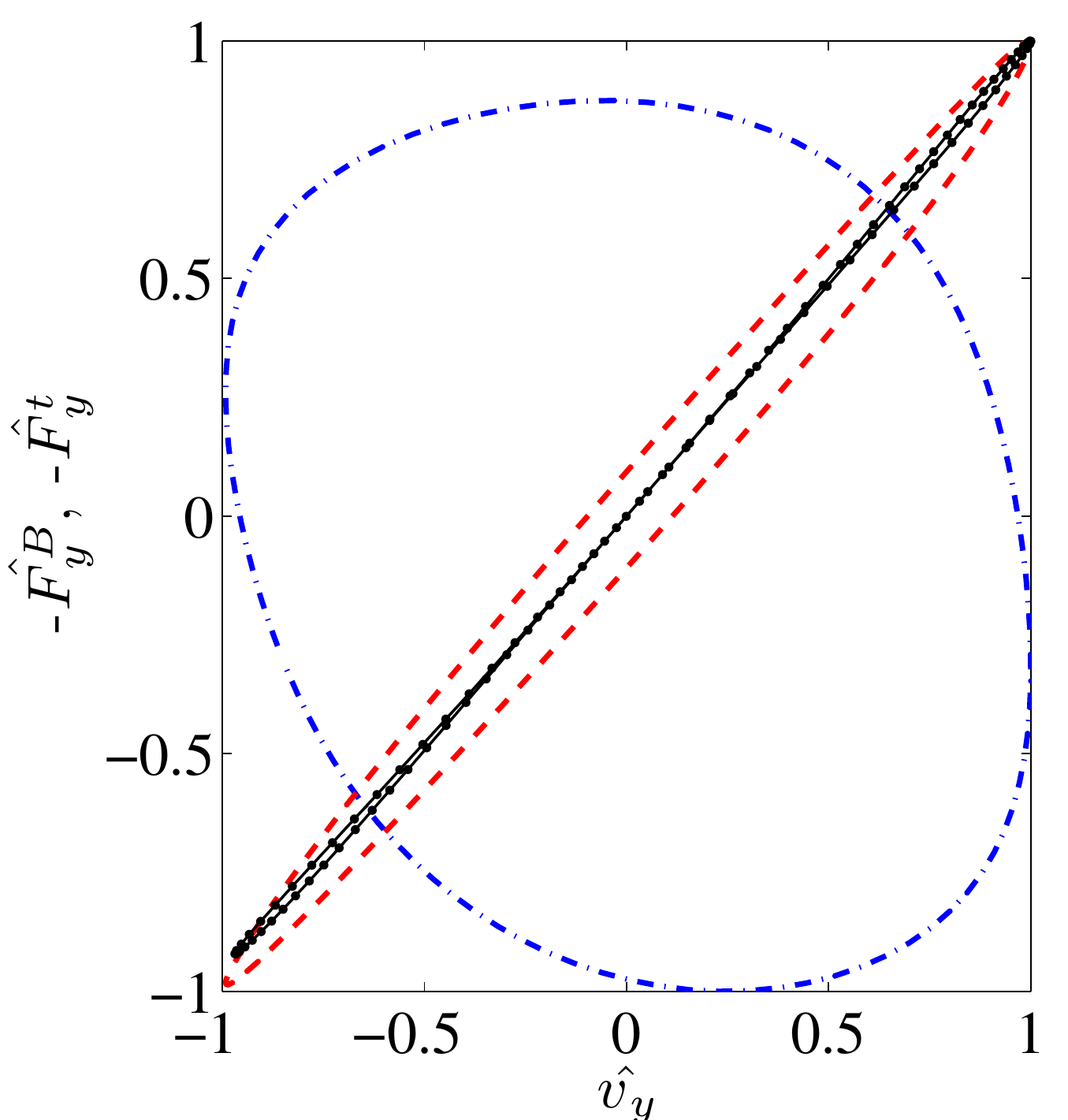}}
\resizebox{4.4cm}{!}{\includegraphics{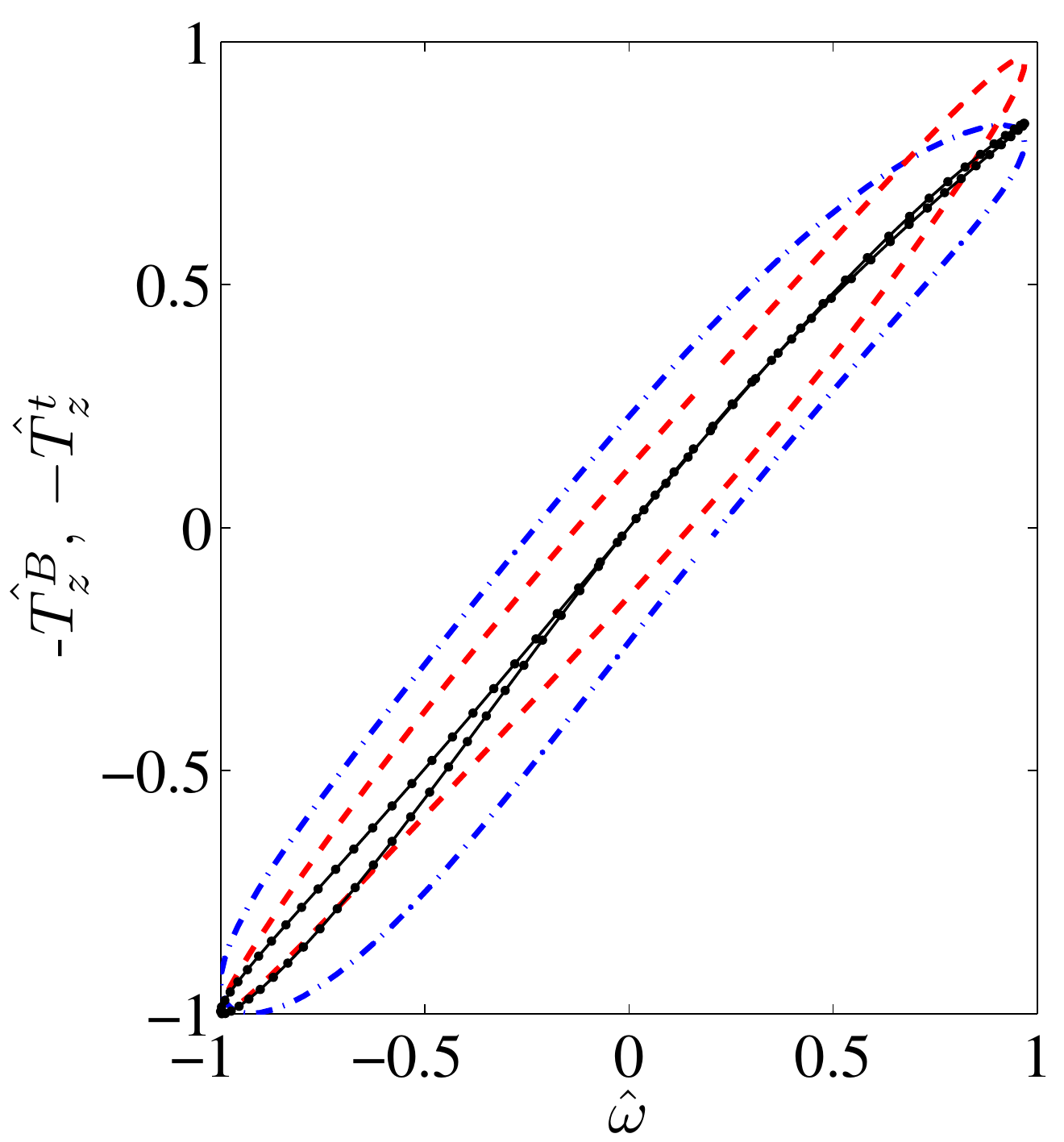}}
\caption{\label{FTt}
{\it (Color  online)} In dashed-red: surface integral on the swimmer flagellum of minus the $x$ and $y$ component of the force $-F_x^t$, $-F_y^t$ (left and middle panel) and $z$ component of the torque $-T_z^t$ (right panel), versus, respectively, the swimmer velocity in $x$, $y$, and the angular velocity. In dotted-dashed blue and dotted black we show similar curves for the integral of the forces on the flagellum due to the beating motion only (right hand side of system (\ref{systemswimming1})-(\ref{systemswimming3})), the dotted black case corresponds to the values in the coordinates of the diagonal system. All the variables in these plots are normalized by their maximum absolute value and for this reason are marked by a hat.}
\end{figure*}

In Fig. \ref{FTt} we plot in dashed-red the total forces and torque exerted by the head to the tail versus the swimmer linear and angular velocities. These results were obtained for the full-body regularized stokeslet model calculation. We find that the points do not lie on a straight line as would be expected for the flow past an isolated sphere or ellipsoid, instead, they trace closed curves. The deviation from a straight line quantifies the contribution due to the head-tail interaction and further demonstrates the limitations of the approximate approach. Lines fitted to the dashed-red curves of Fig. \ref{FTt} have slopes smaller than the theoretical coefficient $6 \pi \mu a$ by an ~8.5\%  in the $x$-direction, larger by  a ~9.8\% in the $y$-direction and smaller than $8 \pi \mu a^3$ by ~6.9\% for the torque in $z$. Similar results hold when comparing the theoretical friction coefficients for prolate ellipsoids with the slopes of lines fitted to analogous curves for swimmers with ellipsoidal heads.


\begin{table*}[htp]
\footnotesize
\caption{Difference between the resistive matrix coefficients and known terms of system (\ref{systemswimming1})-(\ref{systemswimming3}) computed with the simplified assumption of representing the swimmer as the superposition of the head and tail flow field [resistive matrix (\ref{tail+head})] and the full-swimmer representation. The error is computed according to formula (\ref{errorCoeff}). The second row in this table indicates the order of magnitude of the corresponding coefficient. }
\begin{center}
\tabcolsep=0.11cm
\begin{tabular}{ccccccccccccccc}
\hline
$C_x^{v_x}$ 
&$C_x^{v_y}$ 
 &$C_x^{\omega}$ 
 &$C_y^{v_x}$ 
 &$C_y^{v_y}$
 &$C_y^{\omega}$
 &$m_z^{v_x}$ 
 &$m_z^{v_y}$ 
 &$m_z^{\omega}$ 
 &$F_x^B$ 
 &$F_y^B$
 &$T_z^B$ \\
\footnotesize{$\sim$1e-4}
&\footnotesize{$\sim$1e-6}
&\footnotesize{$\sim$1e-10}
&\footnotesize{$\sim$1e-6}
&\footnotesize{$\sim$1e-4}
& \footnotesize{$\sim$1e-9}
&\footnotesize{$\sim$1e-10}
&\footnotesize{$\sim$1e-9}
&\footnotesize{$\sim$1e-14}
&\footnotesize{$\sim$1e-8}
& \footnotesize{$\sim$1e-8}
&\footnotesize{$\sim$1e-12}\\
\hline
23.47 \%            & -4.25\%          & 4.53\%                & -4.25\%        & 17.21\%        &   5.7\%               &   4.53\%        & 5.7\%         & -0.59\% & 7.07\% & 2.9\% & -0.26\% \\
\hline
\end{tabular}
\end{center}
\label{Headon-offTable}
\end{table*}%

\begin{figure} 
\resizebox{6cm}{!}{\includegraphics{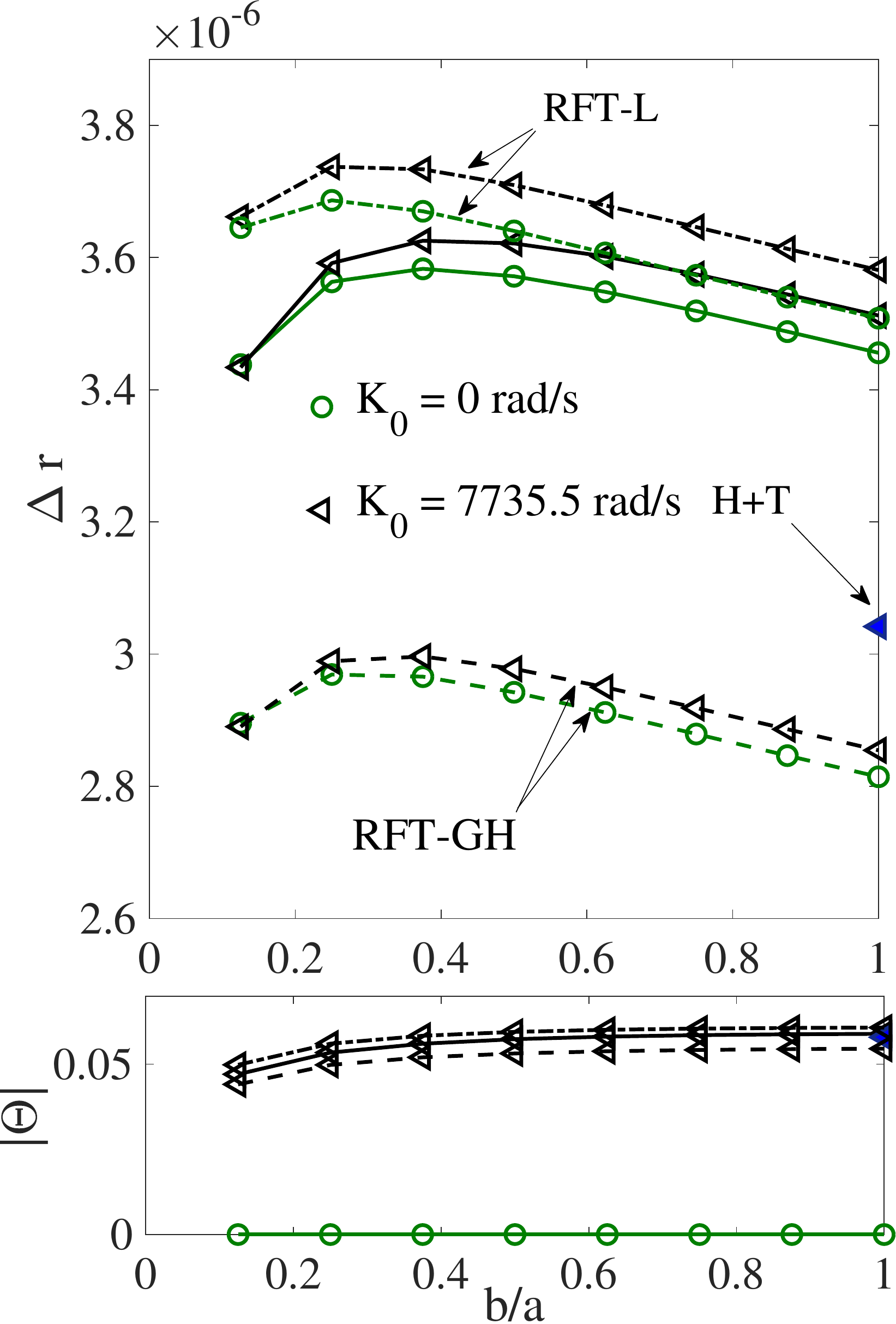}}
\resizebox{5cm}{!}{\includegraphics{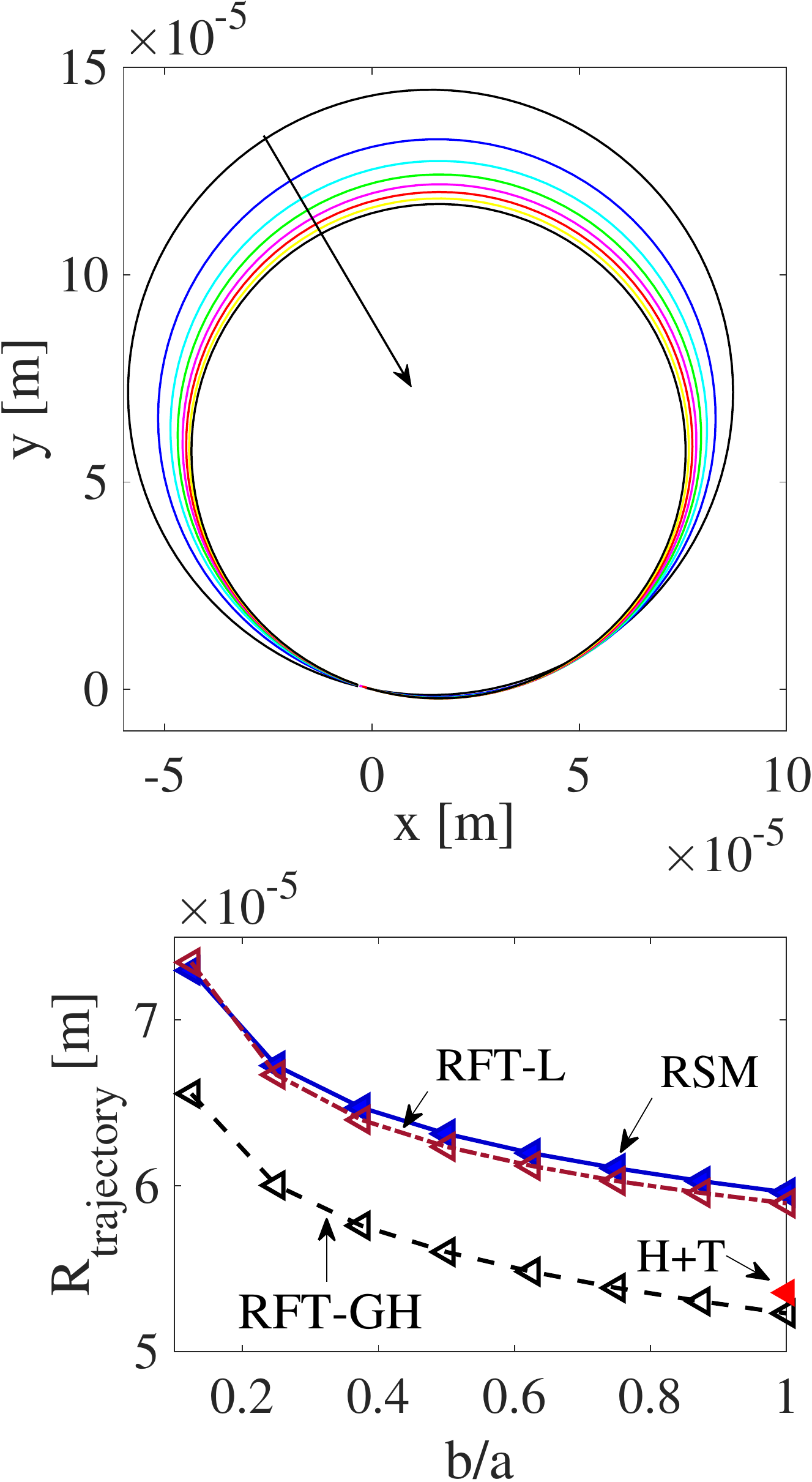}}
\caption{\label{Ellipsoidal}
{\it (Color  online)} {\it LEFT:} Net displacement (top) and rotation angle (bottom) for one beating period as a function of the aspect ratio that characterizes the ellipsoidal head. The curves correspond to the case of null (green circles) and non-null (black triangles) mean flagellar curvature and the Regularized Stokeslet Method (solid line), the RFT Lighthill model (dashed-dotted line), the RFT Gray and Hancock model (dashed-line) and the head+tail model (filled triangle).
{\it RIGHT:} Top: swimming trajectories on the $x$-$y$ plane, drawn by using the values of the net displacement, $\Delta r$, and net rotation angle, $\Theta$, for the swimmer with non-null mean curvature. The arrow indicates the direction of growing $b/a$. Bottom: radius of the trajectory as a function of $b/a$ for the regularized stokeslet model (solid line), the RFT Lighthill model (dash-dotted line) and the RFT Gray and Hancock model (dashed line), the red triangle refers to the head+tail model.}
\end{figure}
\subsection{Sensitivity to the head shape}\label{head}
We study how swimming is affected by the head shape, in particular, we consider the case of swimmers with a prolate spheroidal head. 
We keep the head volume constant while varying the minor to major axis ratio $b/a$. In Fig. \ref{vx}-\ref{trajectory} we report results for ellipsoidal heads with an aspect ratio of 0.5 and 0.25 (dotted red and dashed-dotted black curves). Despite the identical beating movement, the tail integral of forces and torque varies as the head shape varies, given the different swimming velocities and head-tail interaction.

We have already quantified in Fig. \ref{vx_approx} and Table \ref{RFT-table} the error of the RFT in predicting the absolute value of the velocity, here we report, in addition, a slight discrepancy between the RFT and the Regularized Stokeslet Model in identifying the fastest swimmer. 
In Fig. \ref{Ellipsoidal} we plot the net displacement and rotation angle $\Theta$ for one period as a function of the minor to major axis ratio $b/a$. Note that for both values of $K_0$ the swimmer that swims the farthest is the one with $b/a=$[0.375 0.5] for the stokeslet model, $b/a=0.25$ for the RFT models.  
We have estimated the error bar associated to the chosen numerical resolution of RSM to be 0.42\%. This value corresponds to the $x$-velocity component error of the RSM solution for solving the analytical problem reported in Fig. \ref{NumericalErrorHead} at the same numerical resolution as the swimmer problem. The $x$-velocity has been selected as the most sensitive solution component since it corresponds to the maximum discrepancy between the RSM solution and the reference FEM solution (see Table \ref{RFT-table}). The swimmers that possess the largest $v_x$ average velocity according to the RSM are those with $b/a=$[0.625 1], while for the RFT models the maximum $v_x$ average velocity is achieved for $b/a=0.625$.

The results in Fig. \ref{Ellipsoidal} (left) reiterate that for the wavenumber of choice the RFT-L model is more accurate than the RFT-GH model. Note that both the RFT models badly capture the behavior for the most stretched shapes. Observe also that the swimmer with a non-null curvature swims farther than the rectilinear swimmer (covers larger distances in one period).

For the $K_0=3375.5$ rad/s swimmer the net displacement and angle $\Theta$ determine the curved trajectories reported on Fig. \ref{Ellipsoidal} (right, top panel). For a given displacement, larger angles correspond to trajectories with smaller radii, so that swimmers with more elongated heads display trajectories with larger radii. The behavior is monotonic with $b/a$ across all models (Fig. \ref{Ellipsoidal} right, bottom panel), surprisingly, the RFT-L model predicts the trajectory radius even better than the head+tail model despite the lower accuracy in the velocities.

\begin{figure*} \centering
\resizebox{6cm}{!}{\includegraphics{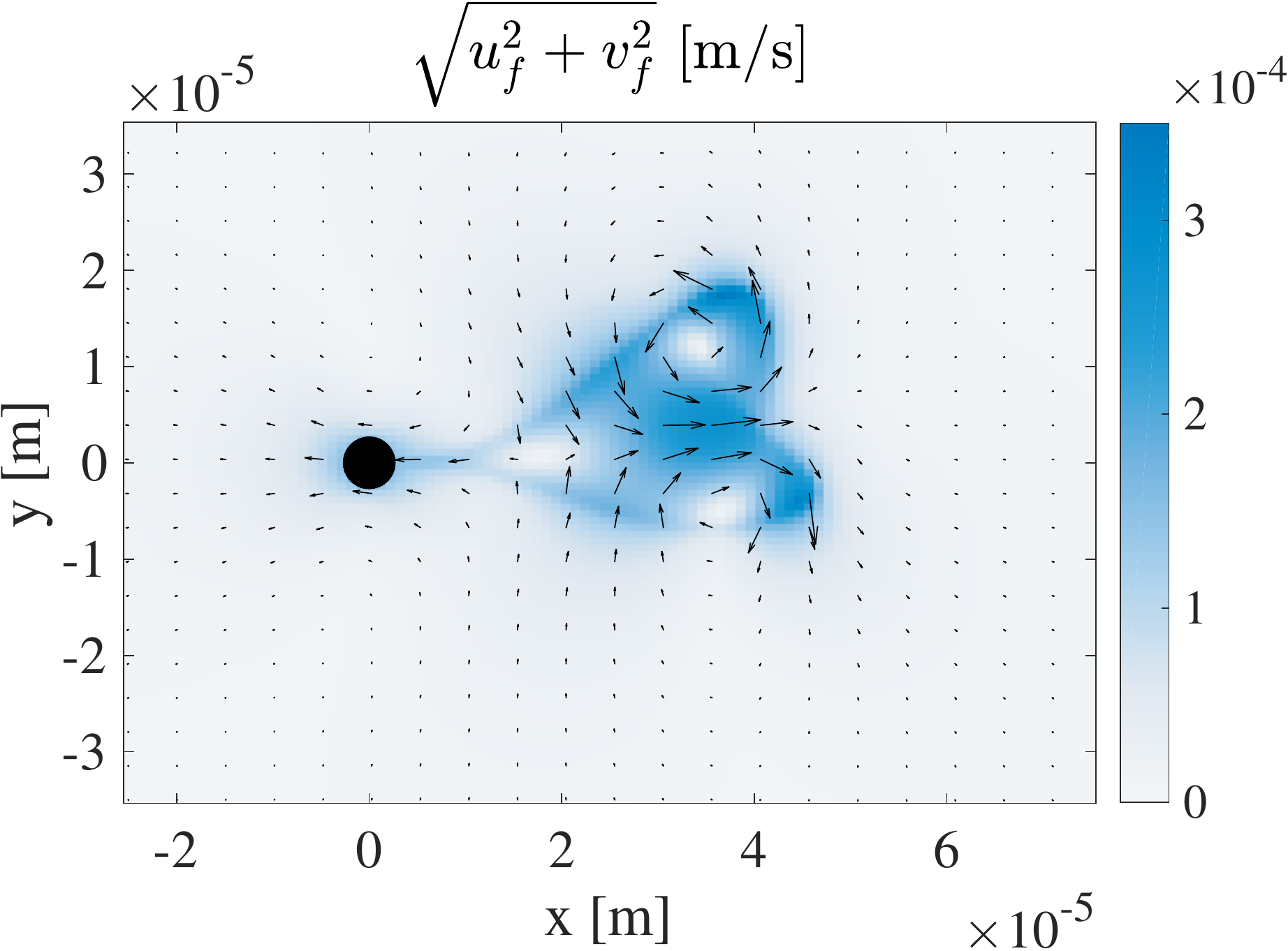}}
\resizebox{6cm}{!}{\includegraphics{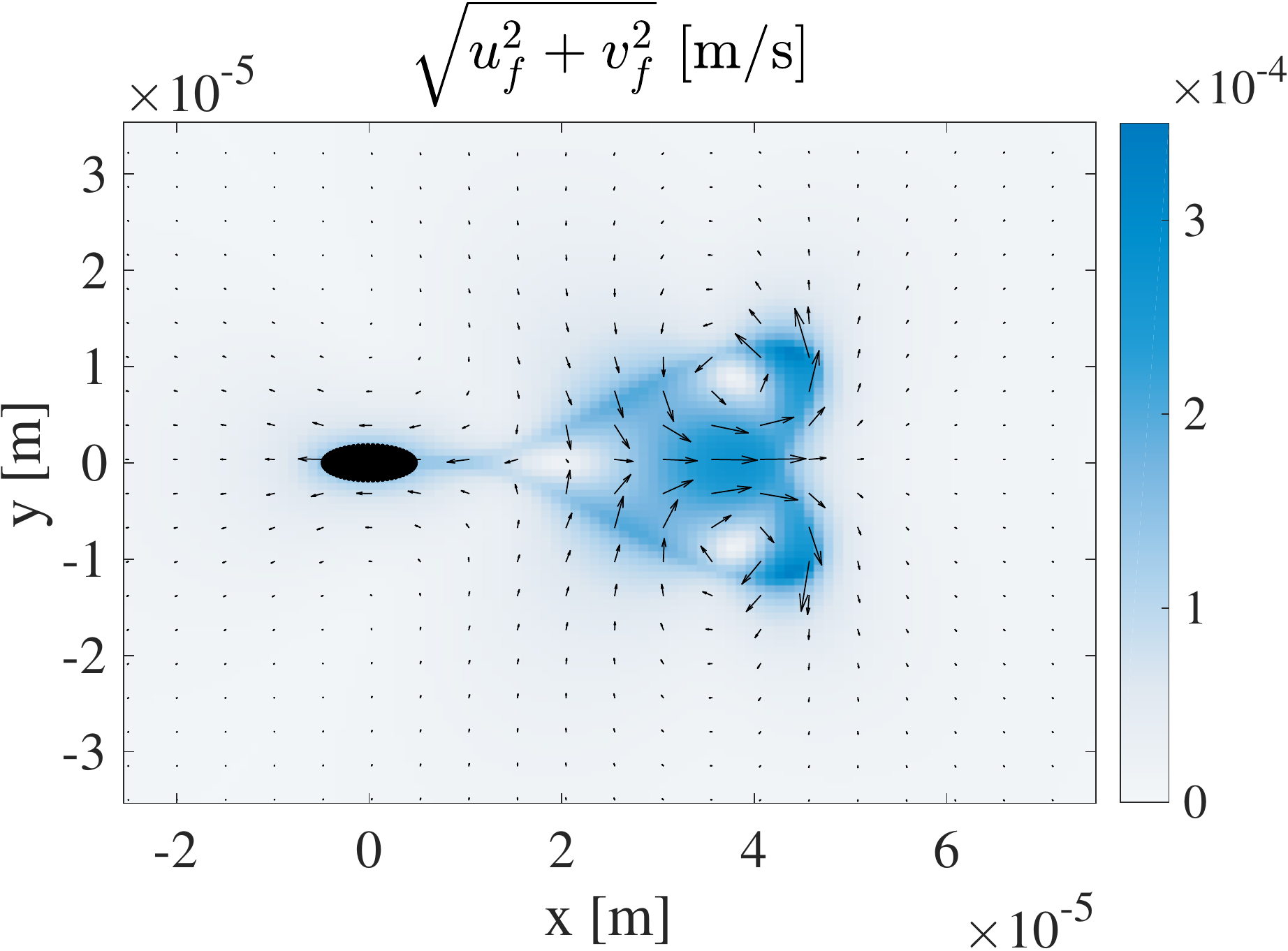}}
\resizebox{6cm}{!}{\includegraphics{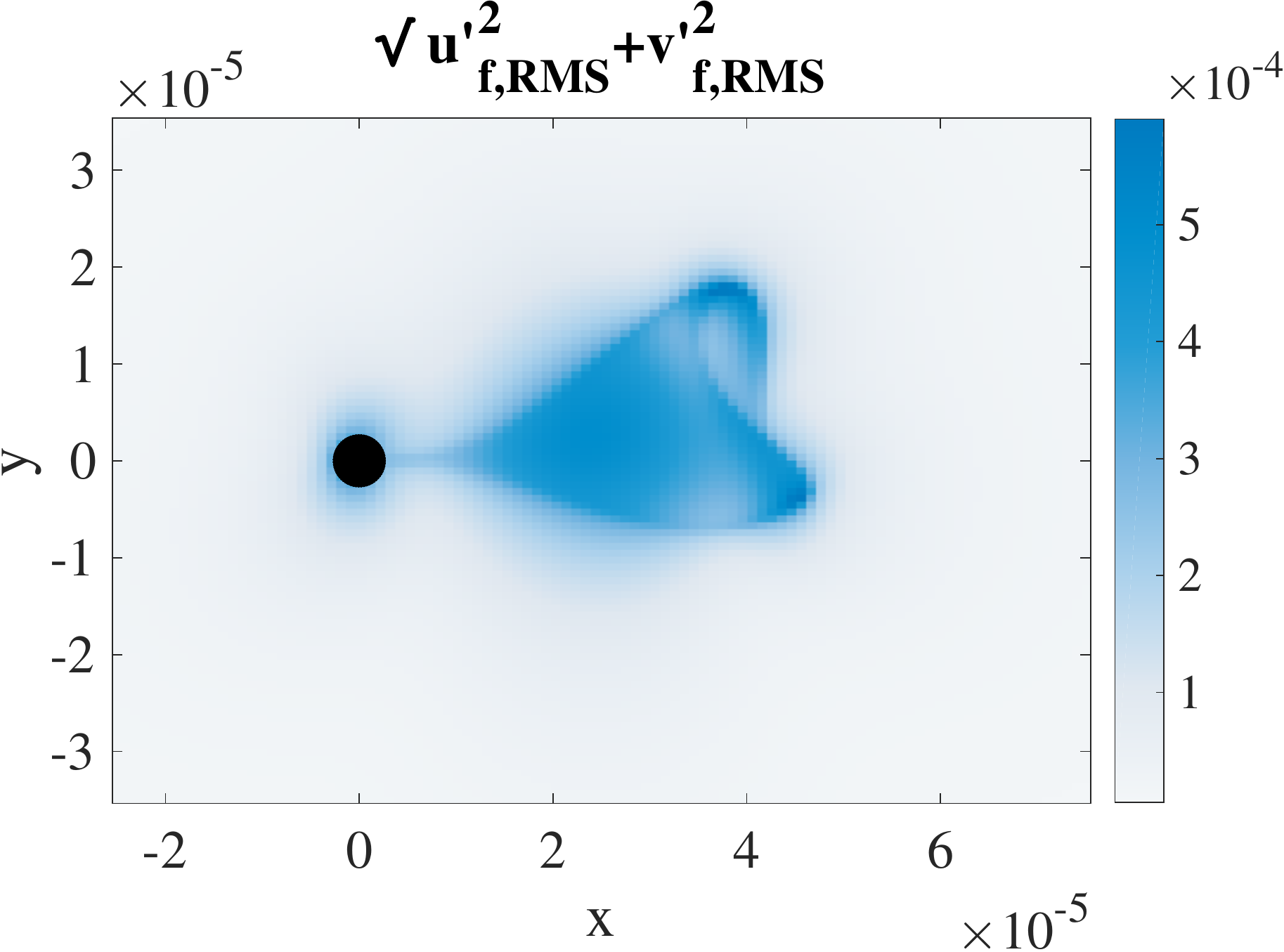}}
\resizebox{6cm}{!}{\includegraphics{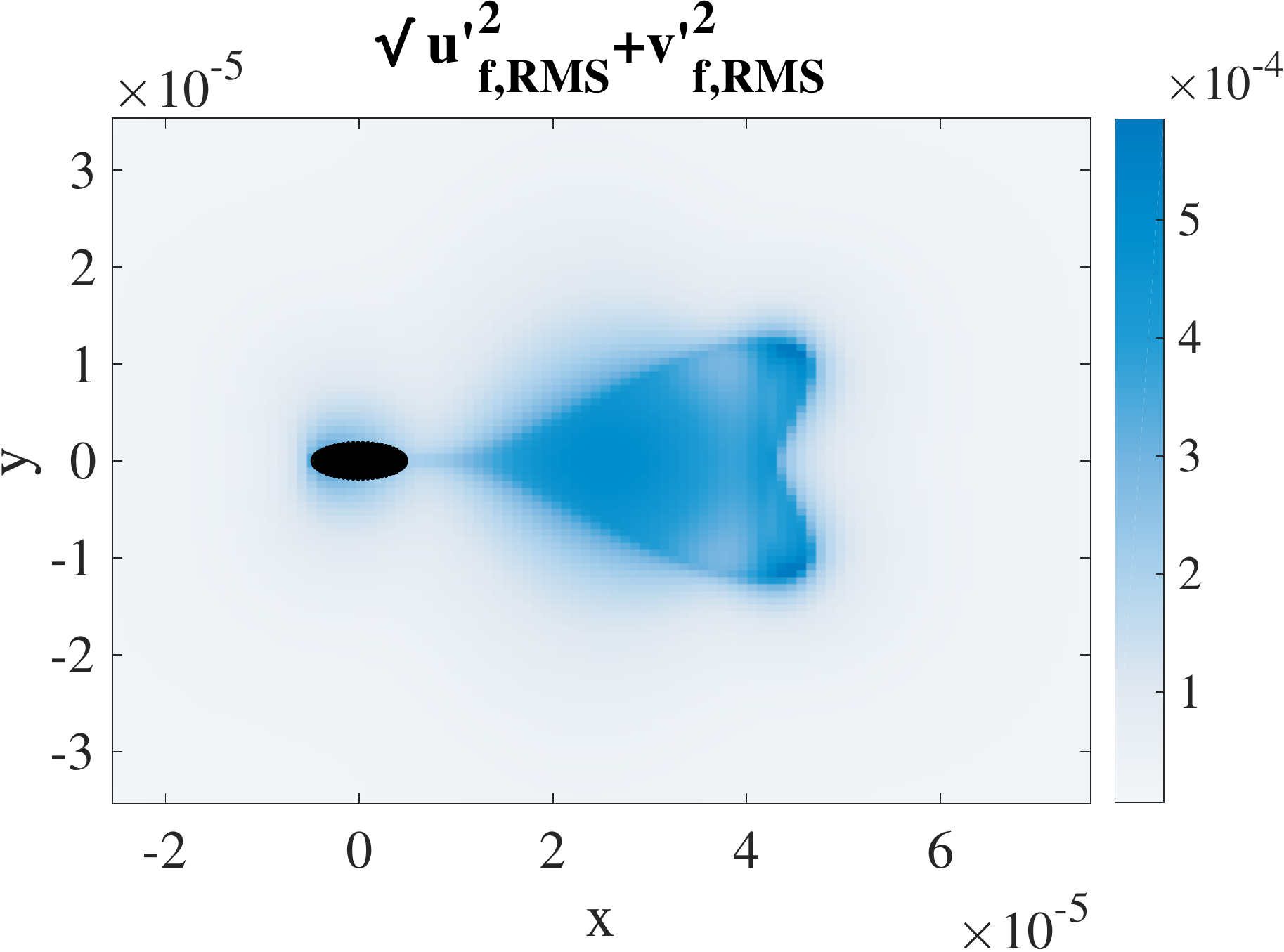}}
\caption{\label{flow_field}
{\it (Color  online)}. {\it Top} - In color: modulus of the average flow field (components $v_f$, $u_f$, $w_f=0$) computed over a beating period in the frame of reference of the swimmer. The black arrows indicate the direction and magnitude of the local field. The left figure represents the case of the swimmer with non-zero curvature $K_0=7735.5$ rad/s, the right figure the calculation for $K_0=0$ and $b/a=0.375$, that is, the fastest swimmer. In black we mark the area occupied by the solid head. {\it Bottom} - modulus of the root mean square value of the velocity fluctuations averaged over one period (eq. \ref{RMSfluctuations}) for the same two cases of the top line.}
\end{figure*}
\begin{figure} \centering
\resizebox{6cm}{!}{\includegraphics{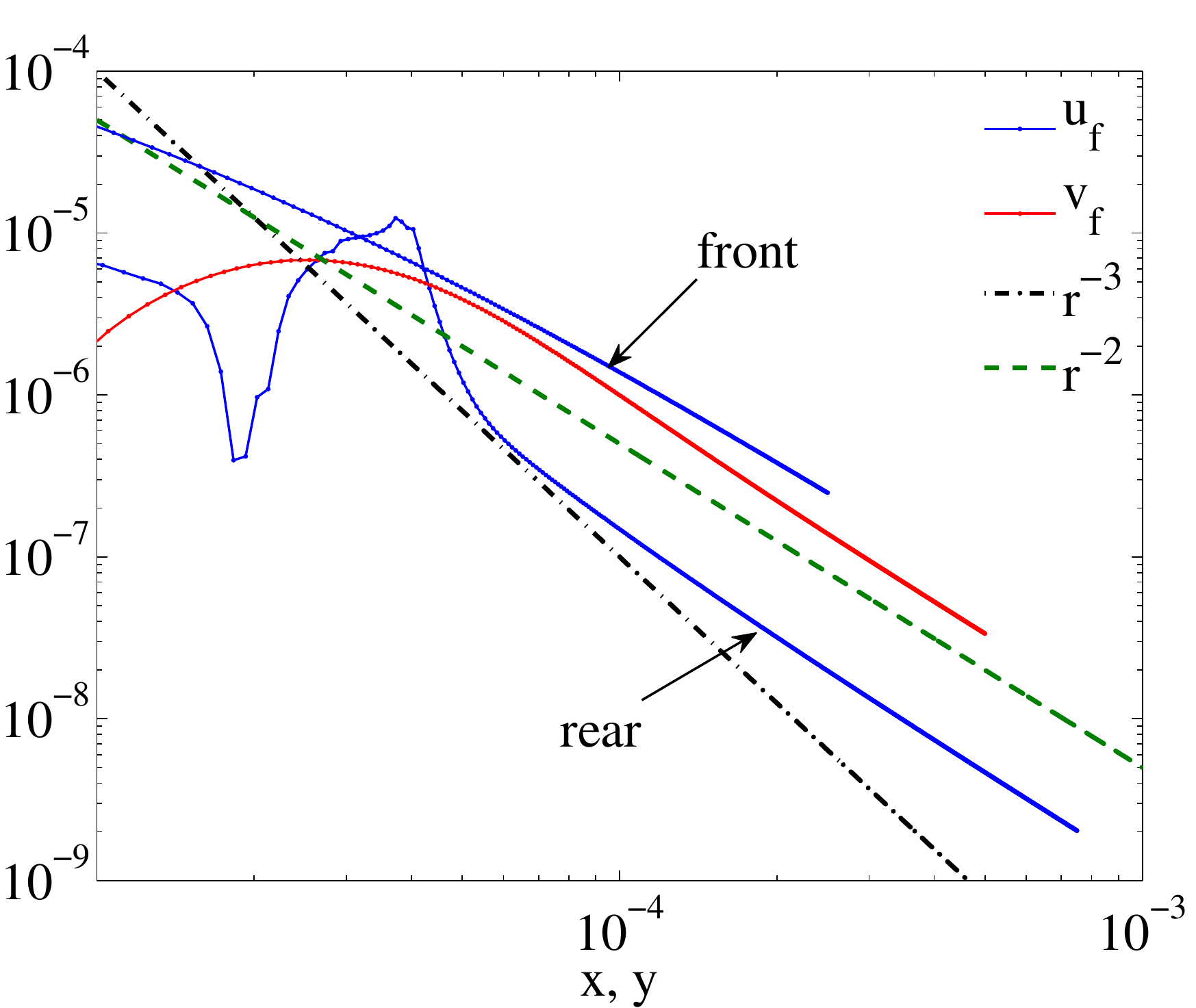}}
\resizebox{5.5cm}{!}{\includegraphics{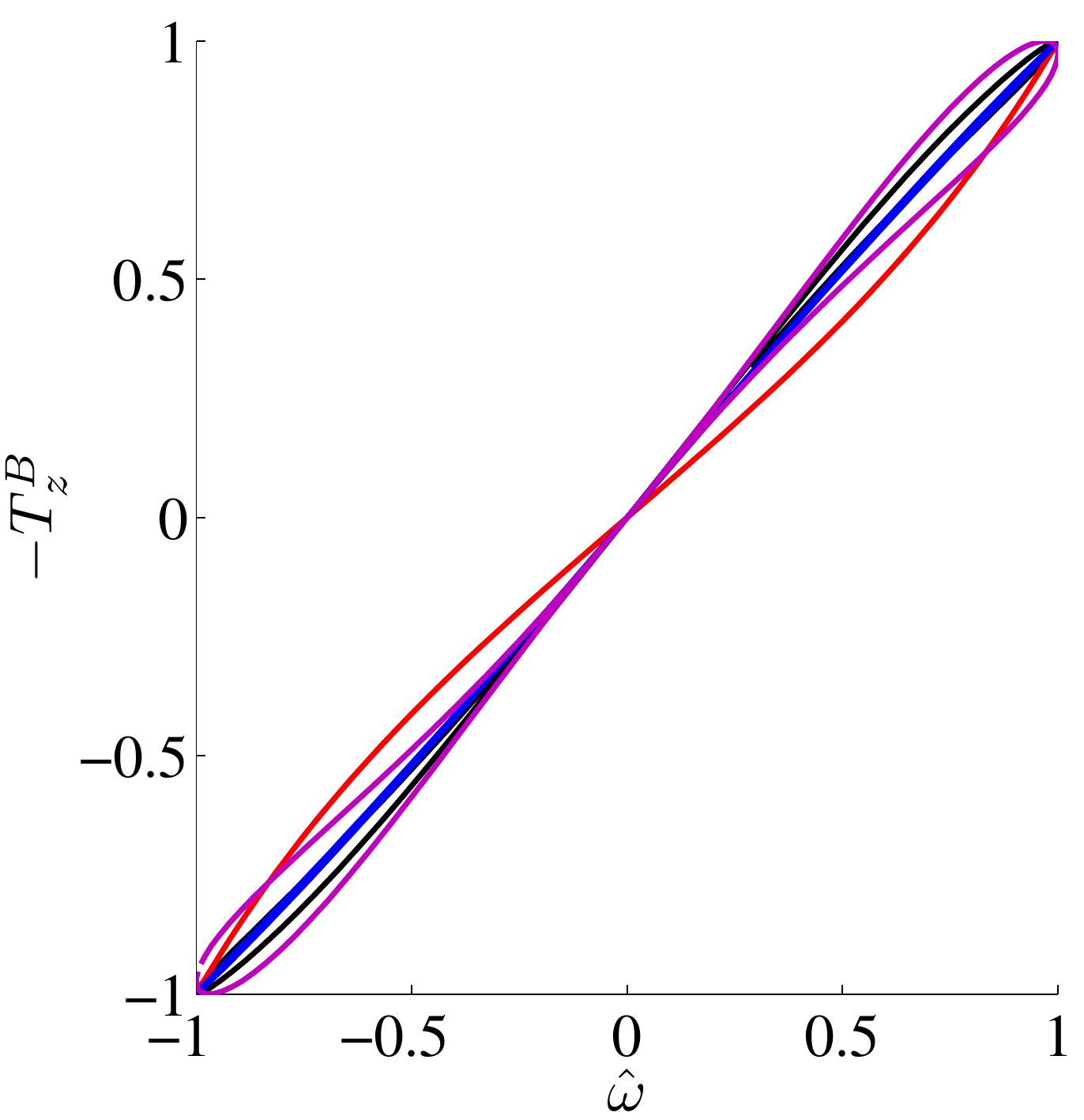}}
\caption{\label{flow_field_prof}
 {\it LEFT:} {\it (Color  online)} Log-log plot of the $u_f$ and $v_f$ average velocity profile respectively along the $y=0$ and $x=0$ cross section on the $z=0$ plane for the $K_0=0$ swimmer with a spherical head. The $u_f$ profile differs on the front and rear of the swimmer as marked by the arrows, while the $v_f$ profile is symmetric with respect to the $y=0$ line. Note the length over which the profile has been computed, one order of magnitude larger than the plots in Fig. \ref{flow_field}, and the $r^{-2}$ far field scaling. {\it RIGHT:} Torque exerted by the flagellum due to the beating motion only versus the swimmer angular velocity for the diagonalized system and the cases: $\lambda=\lambda_0$ (black), $\lambda=\lambda_0/2$ (blue), $\lambda=2\lambda_0$ magenta, $\lambda=100$m (red). All the variables in these plots are normalized by their maximum absolute value.}
\end{figure}

\subsection{Flow field around the swimmer}\label{Flowfield}
In Fig. \ref{flow_field} we show the average flow field $u_f$,$v_f$ and the root mean square velocity fluctuations
\begin{equation}
u'_{f,RMS}=\sqrt{<(u_f-<u_f>)^2>},
\label{RMSfluctuations}
\end{equation}
$$v'_{f,RMS}=\sqrt{<(v_f-<v_f>)^2>},$$
around the swimmer in the frame of reference of the swimmer, here the brackets denote the time average.

We compare the case of swimmers that draw circular and straight trajectories, and for the latter we report results for the swimmer with $b/a=0.375$.  The swimmer flow field resembles the pattern produced by two counterrotating vortex dipoles, one centered at the head, the other centered at about $x=4e-06$. Note that in the left top panel that corresponds to the case of $K_0=7735.5$ rad/s, the second dipole structure is offset with respect to the first one, this reflects the asymmetric beating movement visualized in Fig. \ref{swimmer_shape81} and responsible for the curved trajectory.

In Fig. \ref{flow_field_prof} we display the average flow field profile for a cross section in $x$ and $y$ that spans the interval [0, 5e-04] m, we observe the $r^{-2}$ scaling emerging at large enough distances for respectively the  $u_f$ and $v_f$ component of the velocity. Close to the swimmer the velocity profile does not follow a clear power law and the $u_f$ component dominates.

\subsection{Eigenvalues of the propulsive matrix system}\label{eigenvalues}
The system of equations (\ref{systemswimming1})-(\ref{systemswimming3}) can be diagonalized to remove the effect of the coordinate coupling. When doing so, the curves that on the $v_j-F_j^B$, $\omega_j-T_j^B$ planes draw close loops collapse almost perfectly, and somewhat surprisingly, to a single line as expected for bodies of fixed shape ({\it e.g.} a sphere whose line slope on the $v_j-F_j$ plane would be $6 \pi \mu a$). See Fig. \ref{FTt} and compare the dashed-dotted blue curve versus the black dots curve. The eigenvalues $\Lambda^x(t)$, $\Lambda^y(t)$, $\Lambda^z(t)$ and eigenvectors of the system are in general a function of time, however, the dotted-black curves of Fig. \ref{FTt} can be well fitted by lines $-F_x^B\approx\Lambda^x v_x$, $-F_y^B\approx\Lambda^y v_y$, $-T_z^B\approx\Lambda^z \omega$ of slope $\Lambda_x=0.00011$ N s m$^{-1}$, $\Lambda_y=0.00013$ N s m$^{-1}$, $\Lambda_z=3.7e-14$ N s m. These are not the slopes that appear in Fig. \ref{FTt} where quantities are normalized by their maximum absolute values to allow for comparison. As for the eigenvalues, only the $x$ and $y$ axis appear to change their orientation in time while the $z$ axis along which $\omega$ and $T_z^B$ are directed is fixed.

In conclusion, despite the fact that the swimmer body goes through cyclical deformations, we are able to identify  single time-independent friction coefficients $\Lambda^x, \Lambda^y, \Lambda^z$ that in opportunely rotated frames (the diagonal ones) relate $F_j^B$ to $v_j$, and $\omega$ to $T_z^B$. This fact is consistent with the assumption of a quasi-steady flow described by the time independent Stokes equations. 

However, for larger $\lambda$: $\lambda=2\lambda_0$ and $\lambda=100 $m, the latter representing the case $\lambda\rightarrow\infty$, the friction coefficients of the propulsion matrix show a more marked dependence on time since the points corresponding to the diagonalized system do not lie on a line but draw loops or S-shaped curves in the $\omega_j-T_j^B$ plane, see Fig. \ref{flow_field_prof} (right). Therefore, the flow is sensitive to the change of shape of the beating flagellum and this suggests that the quasi-steady and inertia-less assumptions may break down. It is important to point out that while the frequency Reynolds number is unchanged, the Reynolds number increases as $\lambda$ increases given the larger swimming velocities. The highest Reynolds number for the  $\lambda=100 $m case is $Re=0.035$.


\section{Conclusion}\label{sec:conclusions}
We study numerically the motion of a flagellated microswimmer, specifically a sperm-cell swimmer, in an infinite domain. We first simulate locomotion by using two different numerical methods: the Regularized Stokeslet Method and the Finite Element Method finding a very good agreement between the two.

We find that the Resistive Force Theory performs reasonably well for swimming parameters close to laboratory observations: the normalized root mean square error for the swimming velocities is within 5-15\% depending on the choice of the model parameters. However, the model is unreliable for smaller values of the wavelength, specifically, the case of wavelengths of about 1/4 and 1/2 of the total flagellum length, while the reference case has a wavelength of about one flagellum length. 
These results are consistent with the findings of previous studies that were focused on other types of flagella such as prokaryotic or bacteria-like type \cite{Rodenborn13, Martindale16, Jung07}. For example, in \cite{Rodenborn13} it was reported that RFT fails to provide an accurate description of helical shapes relevant to bacteria, while in \cite{Martindale16} it was concluded that for the broad range of geometry parameters of helical flagellated swimmers considered, RFT never gives accurate results. It was also pointed out that despite being unable to capture the full dynamics, RFT sometimes provides accurate solutions for single quantities \cite{Martindale16}. This is an observation that we also made in reference to the RFT model with the choice of parameters suggested by Lighthill which is able to accurately predict the swimming trajectory radius.
Finally, in \cite{Jung07} it was noted that when studying the complex geometry of superhelices, the experimental results are in excellent agreement with the calculations performed with the RSM, which is not the case of the RFT model.
We also find that the RFT approach fails to correctly predict the optimum shape of the swimmer's head for fastest swimming in the rectilinear case. 

We find that  the simplified approach that consists in studying swimming as the linear superposition of the head and the tail contribution separately, referred to as the head+tail model, leads to inaccurate results and we precisely quantify the error for the velocities, trajectory and force distribution. The inaccuracy of the method originates from the fact that the interaction terms between the head and the flagellum are neglected, or, else, from a physical perspective, the model fails to account for the front-rear symmetry breaking of the flow past the ellipsoidal head due to the presence of the flagellum. For a circular swimmer the radius difference between the trajectory of the head+tail and full-body model is comparable to the radius difference between the trajectory of a swimmer with a spherical head and a swimmer with an ellipsoidal head of minor to major axis ratio ~0.25. This suggests that this simplified approach as well as the RFT is unsuitable to perform optimization studies on hydrodynamically efficient body and beating shapes. 

Finally, we have revealed by diagonalising the propulsion matrix that the flow is rather insensitive to the cyclic deformation of the swimmer body for our choice of parameters $K_0$, $A_0$ and $\lambda_0$. In fact, during one swimming cycle the points that correspond to different instant of time and therefore different shape configurations, lie on an almost perfect straight line on the $v_x$-$F_x^B$, $v_y$-$F_y^B$, and $\omega_z$-$T_z^B$ plane, a behavior typical of fixed-shape objects for which the friction coefficients are given. Consider however, that the friction coefficient is approximately constant provided that the axis are opportunely rotated as the flagellum moves, this is where the analogy with fixed-shape objects ends. We have also observed  that if we fix the frequency Reynolds number and change the wavenumber $\lambda$ of the flagellum traveling wave by choosing larger values, the curves on the  $\omega_z$-$T_z$ plane start displaying a markedly non-linear behavior. This suggests that for these cases the flow "sees" the object deforming and hence could be prone to time and inertia dependent behaviors. This also calls for a more accurate definition of the relevant Reynolds number in case of the flagellated swimmers compared to those 2 commonly used in the literature, which are based on the swimmer length or its beating frequency.







\section*{Acknowledgements}
C.R. thanks Professor Antonio DeSimone, Dr. Vasily Kantsler and Dr. Nicola Giuliani for useful hints and scientific discussions. Part of this work was done while C.R. was visiting SISSA, whose hospitality is gratefully acknowledged. This project has received funding from the European Union's Horizon 2020 research and innovation programme under the Marie Sklodowska-Curie grant agreement No 703526.





%
%
%
%
%

%
%
%
%

\bibliography{CRorai_tex.bbl}
\end{document}